\documentclass[aps,prapplied,reprint,superscriptaddress,longbibliography,floatfix]{revtex4-2}

\usepackage[T1]{fontenc}
\usepackage[utf8]{inputenc}
\usepackage{amsmath,amssymb,bm}
\usepackage{graphicx}
\usepackage{booktabs}
\usepackage{multirow}
\usepackage{xcolor}
\usepackage[hidelinks]{hyperref}

\graphicspath{{figures/}}
\newcommand{\psa}{pSA}
\newcommand{\lpsa}{$\lambda$-pSA}
\newcommand{\tpsa}{$\tau$-pSA}
\newcommand{\sgn}{\operatorname{sgn}}
\newcommand{\E}{\mathbb{E}}

\begin{document}

\title{Error-Rate Reduction in LDPC Decoding via Bit-Aligned Temporal Reinforcement in Parallel Probabilistic-Bit Dynamics}

\author{Naoya Onizawa}
\email{naoya.onizawa.a7@tohoku.ac.jp}
\affiliation{Research Institute of Electrical Communication, Tohoku University, Sendai 980--8577, Japan}
\author{Takahiro Hanyu}
\affiliation{Research Institute of Electrical Communication, Tohoku University, Sendai 980--8577, Japan}

\date{September 8, 2026}

\begin{abstract}
Probabilistic bits (p-bits) provide a physical and algorithmic primitive for stochastic inference, but highly parallel updates can alter their collective dynamics. We study the decoding of random-regular (3, 6) low-density parity-check (LDPC) codes using Jacobi-type p-bit annealing with stochastic partial activation. An additive response-path rule stores each bit's saturated response and reuses it before stochastic readout. High-statistics simulations with independent parameter optimization for each method show pooled bit-error-rate reductions of 33.5\%, 74.8\%, and 81.8\% relative to memoryless probabilistic simulated annealing (pSA) for representative codes of block lengths 96, 192, and 288, respectively. Static gain, normalized averaging, response shuffling, and same-bit binary-state feedback with only its coefficient tuned under the same nonmemory parameters do not reproduce the full saturated-response benefit. Trajectory analysis links the improvement to acquisition of the channel-consistent valid-codeword basin and enhanced post-acquisition stability; the acquisition advantage persists from random and channel-hard-decision starts under the tested conditions. Across all 30 independent code realizations, fixed additive parameter-and-readout packages achieve lower bit- and frame-error rates than separately optimized pSA-specific packages, with neither package retuned for individual codes. These results show that the computational effect of temporal state depends on the retained quantity and its reinjection into stochastic dynamics.
\end{abstract}

\maketitle

\section{Introduction}
\label{sec:introduction}

Probabilistic bits (p-bits) couple a binary stochastic output to a continuously tunable response and provide a hardware-oriented primitive for invertible logic, sampling, and physics-inspired optimization \cite{camsari2017prx,faria2017lmag,borders2019nature}.  Architectures have scaled this concept from stochastic magnetic junctions to autonomous coprocessors and sparse probabilistic Ising machines \cite{sutton2020access,aadit2022natureelectronics,chowdhury2023fullstack}.  Update organization is part of the algorithm: schedules, partial activation, and the degree of simultaneous commitment can change convergence, even when the instantaneous p-bit response law is unchanged \cite{onizawa2023ojsp,onizawa2024srep,onizawa2026landscape}.  Parallel field evaluation is attractive for throughput, but naive simultaneous updates can also produce correlated oscillations.

Temporal state in stochastic Ising dynamics is not new, and different methods retain different variables for different purposes.  Kinetic Ising models have incorporated persistence fields \cite{caccioli2008memory}; momentum annealing and coherent-Ising-machine momentum use configuration or continuous dynamical history \cite{okuyama2019momentum,brown2024momentumcim}.  Spike-frequency adaptation forms a real-valued adaptive state from binary activity and applies negative feedback on the input or energy-gradient side to promote escape from local minima \cite{xu2025sfa}.  PIMI adds the current binary spin after its nonlinear response to stabilize fully synchronous dense updates \cite{zhu2026pimi}, while time-dimensional exchange coupling relates successive binary spin configurations \cite{du2026tec}.  Physical p-bits and coupling paths can also have finite settling and circuit response times \cite{mcgoldrick2022settling,gibeault2024coupling}.  These mechanisms cannot be summarized by a single memory coefficient: stored quantity, sign, scaling, and reinjection all determine the transition kernel.

LDPC decoding provides a structured setting in which parity feasibility and channel evidence must be satisfied together \cite{gallager1962ldpc,mackay1996ldpc,richardson2008coding}.  Statistical-physics treatments established this inference viewpoint \cite{saad2001ldpcphysics}; subsequent Ising and QUBO decoders include early mapped formulations, constraint-aware CMOS annealing, and a relaxation-oscillator solver with native six-body interactions \cite{tawada2020ldpc,elmitwalli2025constraint,dikopoulos2025rxoldpc}.  Most recently, Volpe \textit{et al.} demonstrated p-bit LDPC decoding in a deeply pipelined FPGA architecture \cite{volpe2026pipelined}.  Their pipeline overlaps arithmetic while explicitly preserving sequential logical state-update order.  The present work addresses a different regime: Jacobi-type field evaluation with stochastic partial activation.  Belief propagation remains a conventional decoding reference; p-bit LDPC decoding itself is not the novelty claimed here.

The resulting question is narrower: what form of bit-local temporal state, if any, improves LDPC inference under parallel p-bit dynamics?  The present contribution is not temporal memory or same-bit feedback per se.  We examine a response-path rule in which the bit-specific real-valued saturated response itself is retained and additively reused downstream of the current nonlinear response and before stochastic readout.  Matched gain, normalization, and shuffling controls test amplitude, averaging, and identity; a binary-state self-feedback control with only its feedback coefficient swept tests whether same-bit binary inertia is sufficient.

Within the tested random-regular $(3,6)$ ensemble, additive saturated-response reinforcement promotes acquisition of the channel-consistent valid-codeword basin and improves post-acquisition stability.  The advantage persists under tested alternative initializations and fixed-package transfer to independent matrices.  Binary-state feedback can help, especially at $N=288$, but does not reproduce the full additive benefit.  Separate cross-problem controls reported in the Supplemental Material delimit the scope of generalization.  We therefore claim neither a generic optimizer, a finite-size scaling law, nor a one-to-one mapping between a software coefficient and a device time constant.

\section{P-bit annealing and temporal response dynamics}
\label{sec:dynamics}

\subsection{Factorized objective and instantaneous response}

Let $x_i^{(t)}\in\{-1,+1\}$ denote the signed state of p-bit $i$ at parallel cycle $t$; throughout this paper, binary value $b_i=1$ corresponds to $x_i=+1$.  Each problem is represented by a factorized objective $\mathcal{E}^{(t)}(\bm{x})$ that is linear in any one spin when the others are fixed.  We define the signed local drive by
\begin{equation}
\mathcal{E}^{(t)}(\bm{x})=
\mathcal{E}_{\setminus i}^{(t)}(\bm{x}_{\setminus i})
-x_i F_i^{(t)}(\bm{x}_{\setminus i}),
\qquad
F_i^{(t)}=-\frac{\partial \mathcal{E}^{(t)}}{\partial x_i}.
\label{eq:fielddefinition}
\end{equation}
The derivative denotes the coefficient of $x_i$ in the multilinear extension; it does not restrict the objective to pairwise couplings.  The superscript permits a cycle-dependent stochastic input, used below for the LDPC channel term.  Given the inverse-temperature-like schedule $I_0^{(t)}$, the instantaneous saturated response is
\begin{equation}
q_i^{(t)}=\tanh\!\left[I_0^{(t)}F_i^{(t)}(\bm{x}^{(t)})\right].
\label{eq:q}
\end{equation}
For memoryless \psa{}, the decision variable is $d_i^{(t)}=q_i^{(t)}$.  With independent $\xi_i^{(t)}\sim\mathcal{U}[-1,1]$, its candidate state is
\begin{equation}
\widetilde{x}_i^{(t+1)}=\sgn\!\left(d_i^{(t)}+\xi_i^{(t)}\right).
\label{eq:candidate}
\end{equation}

\subsection{Parallel update with stochastic activation}

The transition kernel can be written as a full candidate step in which every local field and response candidate depends only on the same pre-update state $\bm{x}^{(t)}$ and the same pre-update response state.  State changes are therefore Jacobi, or synchronous, rather than sequential.  Synchronous field dependence does not mean that every p-bit is committed at every cycle.  Independently for each bit and cycle, the implementation draws
\begin{align}
a_i^{(t)}&\sim\operatorname{Bernoulli}(1-p_{\mathrm{hold}}),
\label{eq:activationmask}\\
x_i^{(t+1)}&=
\begin{cases}
\widetilde{x}_i^{(t+1)}, & a_i^{(t)}=1,\\
x_i^{(t)}, & a_i^{(t)}=0.
\end{cases}
\label{eq:activation}
\end{align}
Thus $p_{\mathrm{hold}}$ is the per-bit, per-cycle probability that a p-bit retains its state, and $1-p_{\mathrm{hold}}$ is its activation probability.  The hold probability is fixed within a run and is independent of the annealing schedule.

This stochastic partial activation follows the update-dynamics framework used in our earlier pSA studies \cite{onizawa2024srep,onizawa2026landscape}.  It limits the degree of simultaneous commitment because excessive parallel updating can generate oscillatory behavior and reduce search quality.  Partial activation is common to the algorithms compared here and is not the proposed contribution.  Some parameter sets separately use an initial plateau in a piecewise $I_0^{(t)}$ schedule; this schedule feature is independent of $p_{\mathrm{hold}}$.

\subsection{Additive and finite-response temporal state}

Let $m_i^{(t)}$ be the response state available at the start of cycle $t$, initialized to zero.  Additive \lpsa{} uses
\begin{equation}
d_i^{(t)}=q_i^{(t)}+\lambda m_i^{(t)},
\qquad \lambda\geq0,
\label{eq:additive}
\end{equation}
whereas finite-response \tpsa{} uses the convex update
\begin{equation}
d_i^{(t)}=\rho m_i^{(t)}+(1-\rho)q_i^{(t)},
\qquad 0\leq\rho<1.
\label{eq:filter}
\end{equation}
Equation~(\ref{eq:candidate}) then thresholds $d_i^{(t)}$.  The response state is committed under the same activation mask as the bit state.  Define
\begin{equation}
\widetilde m_i^{(t+1)}=
\begin{cases}
q_i^{(t)}, & \text{instantaneous-response rules},\\
d_i^{(t)}, & \text{finite-response rule},
\end{cases}
\label{eq:responsecandidate}
\end{equation}
and commit
\begin{equation}
m_i^{(t+1)}=a_i^{(t)}\widetilde m_i^{(t+1)}
+[1-a_i^{(t)}]m_i^{(t)}.
\label{eq:responsecommit}
\end{equation}
Consequently, a held p-bit retains both its binary state and its stored response.  When $p_{\mathrm{hold}}=0$, $m_i^{(t)}=q_i^{(t-1)}$ for the additive rule, recovering the familiar one-cycle form.  Because Eq.~(\ref{eq:additive}) spans $[-(1+\lambda),1+\lambda]$, it is reinforcement rather than a convex temporal average.  Setting $\lambda=0$ exactly recovers memoryless \psa{}.

In the ungated limit $p_{\mathrm{hold}}=0$, Eq.~(\ref{eq:filter}) is the sampled form of an ideal single-pole response $\tau\dot m=-m+q$ under piecewise-constant input, with $\rho=\exp(-\Delta t/\tau)$.  With stochastic holding, $\rho$ governs relaxation only on activated updates; we do not assign an effective physical time constant to the gated process or identify $\rho$ with a unique device parameter.  Unlike the additive rule, the filtered state remains in $[-1,1]$ and attenuates new information when $\rho>0$.  Setting $\rho=0$ exactly recovers \psa{} under matched random inputs and activation masks.

\begin{figure*}[t]
\centering
\includegraphics[width=0.96\textwidth]{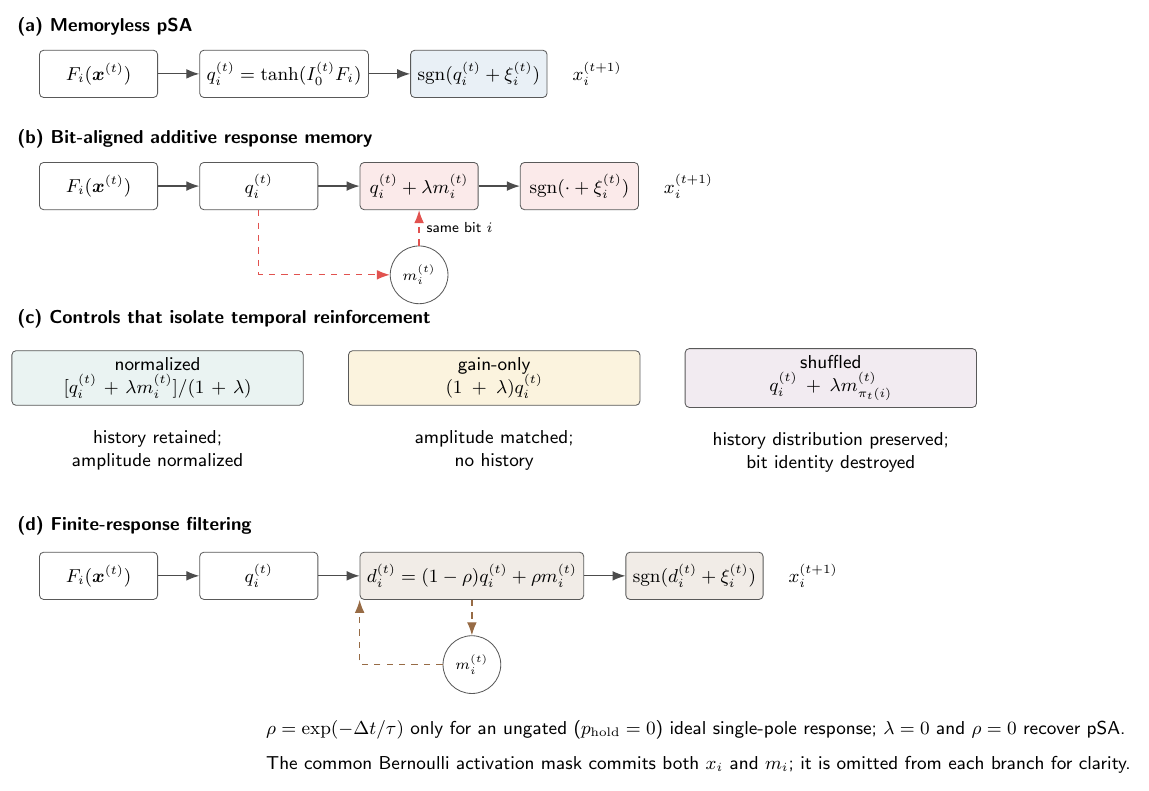}
\caption{Update rules and causal controls.  The additive rule uses the stored response $m_i^{(t)}$ of the same bit, while the shuffled control uses $m_{\pi_t(i)}^{(t)}$ under a cycle-specific derangement $\pi_t$.  Normalized memory preserves temporal state without increasing the deterministic range; gain-only preserves the additive rule's maximum range without temporal state.  The finite-response rule is a convex lag.  All rules use the common Jacobi field evaluation and stochastic activation in Eqs.~(\ref{eq:activation}) and (\ref{eq:responsecommit}).  The diagram distinguishes algorithmic state from a device-specific implementation; no hardware cost or time-constant equivalence is inferred.}
\label{fig:rules}
\end{figure*}

\subsection{Causal controls}

Figure~\ref{fig:rules} places three matched response controls next to the main rule.  Their decision variables are
\begin{align}
d_{i,\mathrm{norm}}^{(t)}&=\frac{q_i^{(t)}+\lambda m_i^{(t)}}{1+\lambda},\\
d_{i,\mathrm{gain}}^{(t)}&=(1+\lambda)q_i^{(t)},\\
d_{i,\mathrm{shuf}}^{(t)}&=q_i^{(t)}+\lambda m_{\pi_t(i)}^{(t)} .
\label{eq:controls}
\end{align}
Here $\pi_t$ is generated as a nonzero cyclic rotation of one randomly ordered bit list, so $\pi_t(i)\neq i$ for every bit while the response multiset, mean, and second moment are preserved.  Normalized memory tests temporal averaging without deterministic-range expansion; gain-only tests the same maximum range without history; and shuffled memory tests whether a distribution-matched but bit-misaligned delayed response is sufficient.

A fourth control replaces the stored response with the same bit's pre-update binary state,
\[
d_{i,\mathrm{bin}}^{(t)}=q_i^{(t)}+\kappa x_i^{(t)},\qquad \kappa\geq0.
\]
We call this \emph{binary-state self-feedback}.  It is inspired by the stored-quantity question raised by PIMI, but it is not a PIMI reproduction: the noise, schedule, stochastic activation, and workload differ.  Unlike Eq.~(\ref{eq:additive}), it retains no separate real-valued response state.  Together, the controls distinguish deterministic range, temporal averaging, bit identity, and the identity of the retained quantity.

Both memory rules require one response state per p-bit and $O(N)$ additional arithmetic.  Sparse local-field evaluation remains $O(|\mathcal{E}|)$ per cycle.  This operation count is not a claim about energy, area, latency, or device efficiency.

\section{LDPC decoding and numerical framework}
\label{sec:framework}

For a parity-check matrix $H\in\{0,1\}^{M\times N}$, a binary estimate is valid when $H\bm b=\bm 0\pmod 2$.  We use $x_i=2b_i-1$, so $b_i=0$ and 1 correspond to $x_i=-1$ and $+1$, respectively.  Every tested random-regular $(3,6)$ matrix has even check degree six; hence check $a$ is satisfied exactly when $\prod_{j\in\partial a}x_j=+1$.

Transmission uses BPSK, $s_i^{\mathrm{tx}}=1-2b_i=-x_i$, over an additive-white-Gaussian-noise channel.  From the received sample $y_i$, the implementation forms $z_i=\tanh[g_i(y_i)]$, where the selected scaling $g_i$ is recorded with each parameter set.  At every cycle it draws $u_i^{(t)}\sim\mathcal U[-1,1]$, sets $r_i^{(t)}=\mathbf 1[z_i<u_i^{(t)}]$, and uses the signed stochastic channel bit $c_i^{(t)}=2r_i^{(t)}-1$.  Thus $\Pr(c_i^{(t)}=+1)=(1-z_i)/2$ and $\E[c_i^{(t)}\mid y_i]=-z_i$.

Conditioned on the cycle's stochastic channel bits, the implemented factorized energy and local drive are
\begin{align}
\mathcal E_{\mathrm{LDPC}}^{(t)}(\bm x)
&=-k_w\sum_a\prod_{j\in\partial a}x_j
-k_r\sum_i c_i^{(t)}x_i, \label{eq:ldpcenergy}\\
F_{i,\mathrm{LDPC}}^{(t)}
&=k_w\sum_{a\in\partial i}\prod_{j\in\partial a\setminus i}x_j
+k_r c_i^{(t)} .
\label{eq:ldpcfield}
\end{align}
The parity term rewards zero syndrome.  In expectation, the channel drive is $-k_r z_i$ and therefore favors the BPSK-consistent bit value.  Parity validity alone is insufficient: a different codeword can also have zero syndrome, whereas the channel term distinguishes the transmitted word among valid codewords.  This is the mathematical basis for the channel-consistent valid-basin language used below.  Detailed sign derivations are given in Appendix~\ref{app:mapping}, and readout rules in Appendix~\ref{app:parameters}.

The primary endpoints are bit-error rate (BER) and frame-error rate (FER) relative to the transmitted word at $E_b/N_0\in\{2.0,2.5,3.0\}$ dB.  Belief propagation (BP) is included in Fig.~\ref{fig:representative} as a conventional decoding reference; no computational, latency, energy, or implementation superiority over BP is claimed.  BP implementation details are given in the Supplemental Material.  All main LDPC studies use $N=96,192,288$ and $M=N/2$.  ``Representative'' denotes one fixed matrix per size, whereas ``independent code realizations'' denotes the distinct matrices in the transfer study and does not imply finite-size scaling.

\subsection{Numerical-study designs}

The LDPC studies answer distinct questions and are not pooled as one comparison.  Table~\ref{tab:protocols} summarizes each design, its parameter relation, and the inference it supports.  Benchmark and control trajectories begin from the implementation's all-zero bit state with zero response state; the correct-start intervention instead begins at the transmitted codeword while retaining zero response state.  Depending on the recorded parameter package, readout uses the final state, a post-burn-in bitwise majority, or the visited state with minimum syndrome weight and maximum channel score to break ties within the readout window (Appendix~\ref{app:parameters}).  Detailed protocols, summary results, and the separate cross-problem studies are provided in the Supplemental Material \cite{supplement}.  Complete machine-readable parameters, seeds, matrix identifiers, validation hashes, and condition-level records are available in the public repository and archive \cite{onizawa_psa_ldpc_zenodo_2026}.

\begin{table*}[t]
\caption{Distinct numerical-study designs.  The comparison unit in the last column is also the resampling unit for the stated cross-instance uncertainty.}
\label{tab:protocols}
\scriptsize
\begin{tabular}{@{}p{0.14\textwidth}p{0.175\textwidth}p{0.225\textwidth}p{0.255\textwidth}p{0.095\textwidth}@{}}
\toprule
\raggedright Study & \raggedright Scope & \raggedright Parameter relation & \raggedright Supported inference & \raggedright Statistical unit \tabularnewline
\midrule
\raggedright Optimized LDPC benchmark & \raggedright One fixed $H$ per $N=96,192,288$; three SNRs & \raggedright Independent tuning by mode; 10,000 trials/SNR/mode & \raggedright Best-achieved performance for the representative matrices & \raggedright Trial; seed batch for paired checks \tabularnewline
\raggedright Causal and stored-quantity controls & \raggedright Representative matrices; high-statistics endpoints and targeted trajectories & \raggedright Matched rule controls plus a separately swept binary-state coefficient; paired streams & \raggedright Rule discrimination, response alignment, and stored-quantity sufficiency & \raggedright Seed batch or trajectory \tabularnewline
\raggedright Acquisition--stability analysis & \raggedright Representative $N=192,288$ matrices; 2.5 dB; 200 trajectories per cell & \raggedright Fixed matched-control sets; all-zero, random, and channel-hard starts; separate correct-start intervention & \raggedright Correct-basin acquisition, initialization robustness, and conditional stability & \raggedright Trajectory or paired seed batch \tabularnewline
\raggedright Cross-code transfer robustness & \raggedright Ten independent $(3,6)$ matrices/size; 1000 trials/SNR/arm/code & \raggedright Fixed independently selected additive and pSA-specific packages; matched $\lambda=0$ rule ablation; no per-code tuning & \raggedright Package-level fixed-transfer comparison and separate response-rule ablation & \raggedright Code realization \tabularnewline
\bottomrule
\end{tabular}
\end{table*}

\section{Representative LDPC performance}
\label{sec:performance}

\begin{figure*}[t]
\centering
\includegraphics[width=0.96\textwidth]{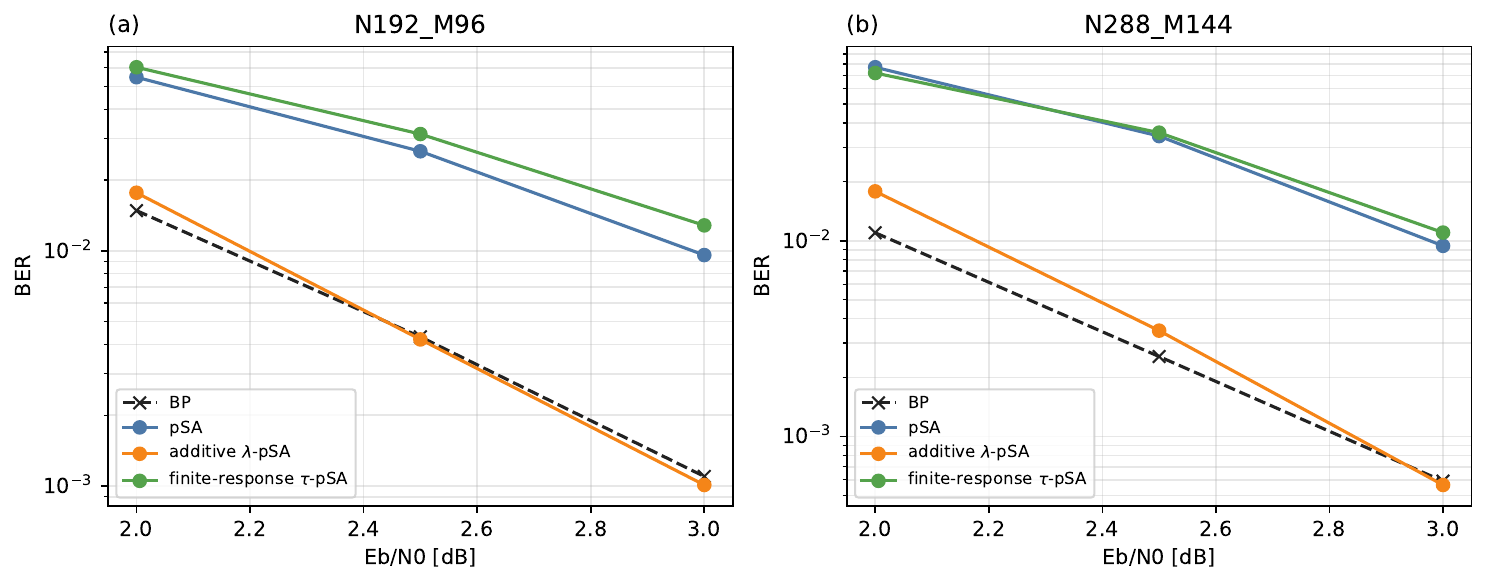}\\[2mm]
\includegraphics[width=0.96\textwidth]{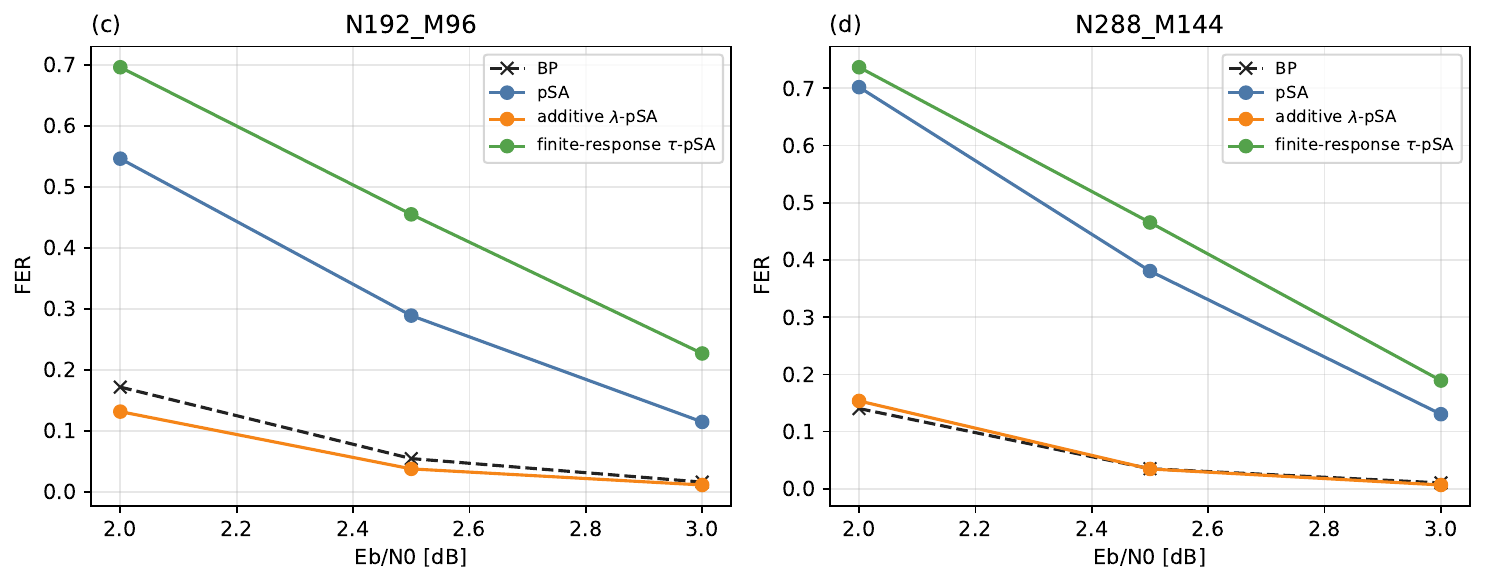}
\caption{Representative-matrix LDPC performance for $N=192$ and 288.  The three p-bit modes were searched and refined independently and use 10,000 trials per SNR; BP uses 5000 trials per SNR in a separate, nonpaired Monte Carlo stream and does not undergo that search.  Coefficient-grid selection and evaluation streams are detailed in Supplemental Sec.~S2.  These are best-achieved benchmark results, distinct from the matched controls in Fig.~\ref{fig:controls} and the fixed-transfer comparisons in Fig.~\ref{fig:crosscode}.  (a,b) BER and (c,d) FER.  The $N=96$ data follow the same protocol and are reported numerically in Table~\ref{tab:representative} and in the Supplemental Material.}
\label{fig:representative}
\end{figure*}

Figure~\ref{fig:representative} compares independently optimized p-bit modes on the representative matrices, using 10,000 trials per SNR and mode.  BP instead uses 5000 trials per SNR from separate Monte Carlo streams and is not paired trial by trial with the p-bit modes; it is not subjected to their parameter search.  The displayed additive and finite-response endpoints are the selected coefficient-grid evaluations, rather than additional held-out estimates; selection and evaluation streams are detailed in Sec.~S2 of the Supplemental Material.  Averaged equally over the three SNRs, additive \lpsa{} reduces BER relative to independently optimized \psa{} by 33.5\%, 74.8\%, and 81.8\% at $N=96,192,288$, respectively (Table~\ref{tab:representative}).  The corresponding pooled FER reductions are 29.2\%, 81.0\%, and 83.9\%.  Finite-response \tpsa{} does not provide a consistent LDPC advantage; its pooled BER is worse than \psa{} for $N=96$ and 192 and only 1.7\% lower at $N=288$.

\begin{table}[t]
\caption{Representative-code results pooled equally over 2.0, 2.5, and 3.0 dB.  Modes were independently optimized.}
\label{tab:representative}
\begin{ruledtabular}
\begin{tabular}{cccccc}
$N$ & \multicolumn{2}{c}{BER} & $\Delta$BER & \multicolumn{2}{c}{FER} \\
 & \psa{} & Add. & (\%) & \psa{} & Add. \\
\hline
96  & 0.02275 & 0.01513 & 33.5 & 0.2616 & 0.1852 \\
192 & 0.03025 & 0.007619 & 74.8 & 0.3165 & 0.0600 \\
288 & 0.04025 & 0.007308 & 81.8 & 0.4043 & 0.0649 \\
\end{tabular}
\end{ruledtabular}
\end{table}

These values establish performance only for the fixed matrices and their separately selected parameter sets.  They do not by themselves identify the responsible dynamical ingredient or establish transfer to new codes.  Matched controls, trajectory analysis, and cross-code tests address those questions.

\section{Causal controls of additive temporal reinforcement}
\label{sec:controls}

\begin{figure*}[t]
\centering
\includegraphics[width=0.98\textwidth]{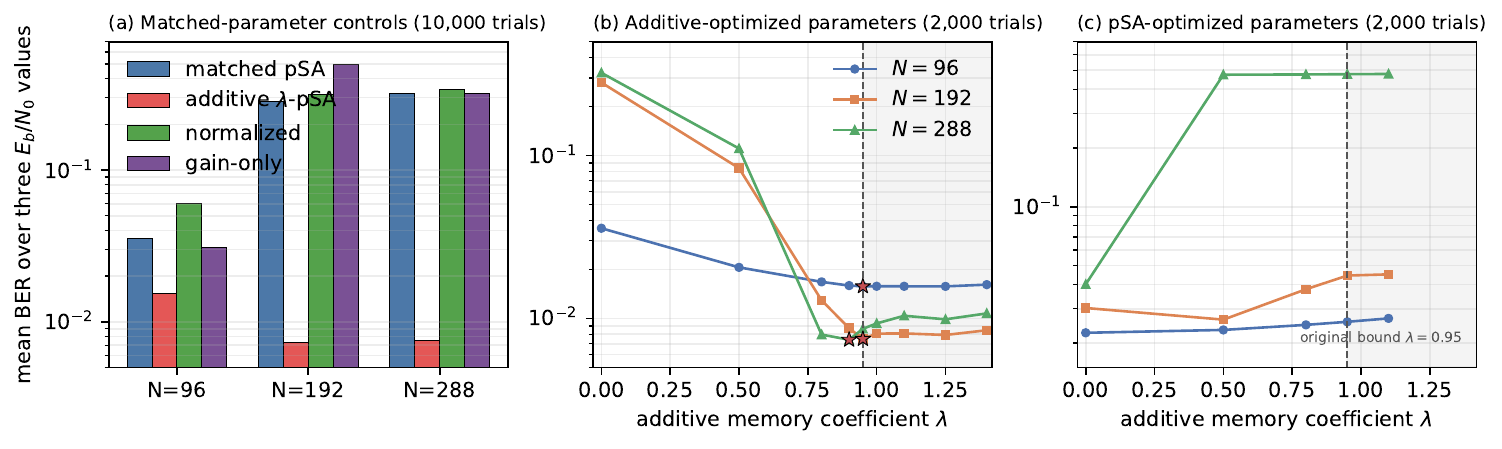}
\caption{Controls and coefficient sweeps on the representative matrices.  (a) Matched controls with fixed additive-optimized nonmemory parameters, 10,000 trials per SNR, and ten shared seed batches; only the response rule changes.  Reported paired intervals for this comparison use seed batches, not individual bits.  (b,c) $\lambda$ sweeps with 2000 trials per SNR and eight shared seed streams within each sweep: (b) additive-optimized and (c) pSA-optimized nonmemory parameters.  The plotted summaries pool equally over the three SNRs; no error bars are drawn here.  Additive response memory outperforms the matched controls in (a).  The coefficient sweeps show no continuing improvement beyond the selected high-$\lambda$ region, so the result is not explained by truncation at the original search limit.}
\label{fig:controls}
\end{figure*}

In the matched causal-control study, the additive-optimized nonmemory parameter set is fixed and only the response rule is varied.  The high-statistics evaluation uses 10,000 trials per SNR and condition with paired seed batches.  The selected coefficients are $\lambda=0.95$, 0.95, and 0.90 for $N=96,192,288$.  Additive BERs pooled over SNR are 0.01540, 0.007365, and 0.007597.  Neither normalized memory nor gain-only approaches these values (Fig.~\ref{fig:controls}).  Against additive, the paired excess BERs and 95\% seed-batch intervals are, respectively, $0.04497$ $[0.04449,0.04544]$ and $0.01573$ $[0.01511,0.01635]$ for $N=96$; $0.30776$ $[0.30720,0.30833]$ and $0.49233$ $[0.49169,0.49297]$ for $N=192$; and $0.33068$ $[0.33026,0.33110]$ and $0.31471$ $[0.31427,0.31515]$ for $N=288$.

The response-alignment analysis adds the shuffled-memory intervention.  The derangement preserves the stored response's mean and second moment to numerical precision but makes the response used by bit $i$ nearly uncorrelated with that bit's own preceding response.  At 2.5 dB, the decoded BERs for additive, matched \psa{}, and shuffled memory are 0.0158, 0.0379, and 0.0958 at $N=96$; 0.00737, 0.2817, and 0.3837 at $N=192$; and 0.00523, 0.3201, and 0.3965 at $N=288$.  The benefit therefore requires bit-specific temporal alignment, not merely the marginal distribution of a delayed signal.

Binary-state self-feedback addresses a different estimand, so it is not added to Fig.~\ref{fig:controls}.  Its coefficient $\kappa$ is swept separately to give binary feedback a best-performance comparison rather than forcing $\kappa=\lambda$.  At $N=96$ and 288 the score-, BER-, and FER-selected values coincide at $\kappa=0.25$ and 0.75.  Their pooled BERs are 0.03029 and 0.01952, compared with additive values 0.01540 and 0.00760; binary-minus-additive paired differences are 0.014883 $[0.014209,0.015558]$ and 0.011927 $[0.011260,0.012593]$.  At $N=192$, the score-selected $\kappa=0.50$ freezes in a wrong valid codeword.  A separately high-statistics, BER-favorable sensitivity at $\kappa=0.25$ still has BER 0.21152, exceeding additive by 0.204155 $[0.203056,0.205253]$.  Equal-range $\kappa=\lambda$ tests and the complete selection audit are reported separately in the Supplemental Material \cite{supplement}.

Together these interventions reject deterministic gain alone, normalized averaging alone, distribution-matched but misaligned response history, and same-bit binary inertia as complete explanations.  Binary feedback is not generally ineffective: at $N=288$ it captures a substantial part of the improvement over matched \psa{}, but not the full saturated-response benefit.

\section{Dynamical mechanism in LDPC decoding}
\label{sec:mechanism}

\subsection{Acquisition from the default initialization}

The acquisition--stability analysis follows trajectories initialized from the implementation's default all-zero bit state at 2.5 dB, with zero response state.  Because every linear code contains the all-zero codeword while the transmitted codeword is randomly generated, this initial state is valid but is generally not the transmitted word.  A trajectory is said to acquire the correct basin on its first visit to the transmitted valid codeword.  This targeted analysis compares matched \psa{}, additive, shuffled-memory, and gain-only dynamics; normalized memory is evaluated in the matched causal-control study of Sec.~\ref{sec:controls} but is not part of this acquisition--stability data set.  For $N=192$ and 288, additive dynamics reach the correct state in 95.5\% and 95.0\% of 200 trajectories, whereas matched \psa{}, shuffled memory, and gain-only reach it in 0\% under this protocol.  Among additive trajectories that reach, the median first-passage cycles are 444 and 425.  Thus the principal inter-method difference under the default initialization is acquisition, not a comparison conditioned on all methods first reaching the target.

To test whether this difference is specific to initialization at a valid all-zero codeword, we repeated the matched-parameter analysis from independent random spins and channel-hard-decision states, again using 200 trajectories per method, initialization, and size.  At $N=192$, additive dynamics acquire the transmitted codeword in 96\% of trajectories from either alternative start; at $N=288$, the corresponding rates are 96\% and 97\%.  Matched \psa{}, gain-only, and shuffled memory reach it in 0\% in all four alternative-initialization cells.  The additive decoded BER remains 0.0035--0.0059, compared with 0.28--0.32 for matched \psa{}.  The acquisition advantage is therefore not an artifact of initialization at the valid all-zero codeword under these tested conditions; complete initial-state diagnostics and censored first-passage curves are reported in the Supplemental Material \cite{supplement}.

Figure~\ref{fig:mechanism} links this outcome to channel consistency, parity feasibility, and state motion.  Additive trajectories jointly reduce syndrome and instantaneous BER while aligning with the channel evidence.  At 2.5 dB the late correct-to-incorrect back-flip rates are $3.63\times10^{-4}$ and $1.70\times10^{-4}$ for $N=192$ and 288, compared with 0.144 and 0.175 for matched \psa{}; the denominator includes all previously correct bit states, including held bits (Appendix~\ref{app:metrics}).  The retained-response/selected-source product remains about 0.996 for bit-aligned additive memory but collapses toward zero after shuffling.  This uncentered diagnostic includes held cycles and is defined separately from same-bit response persistence in Supplemental Sec.~S5.

\begin{figure*}[t]
\centering
\includegraphics[width=0.45\textwidth]{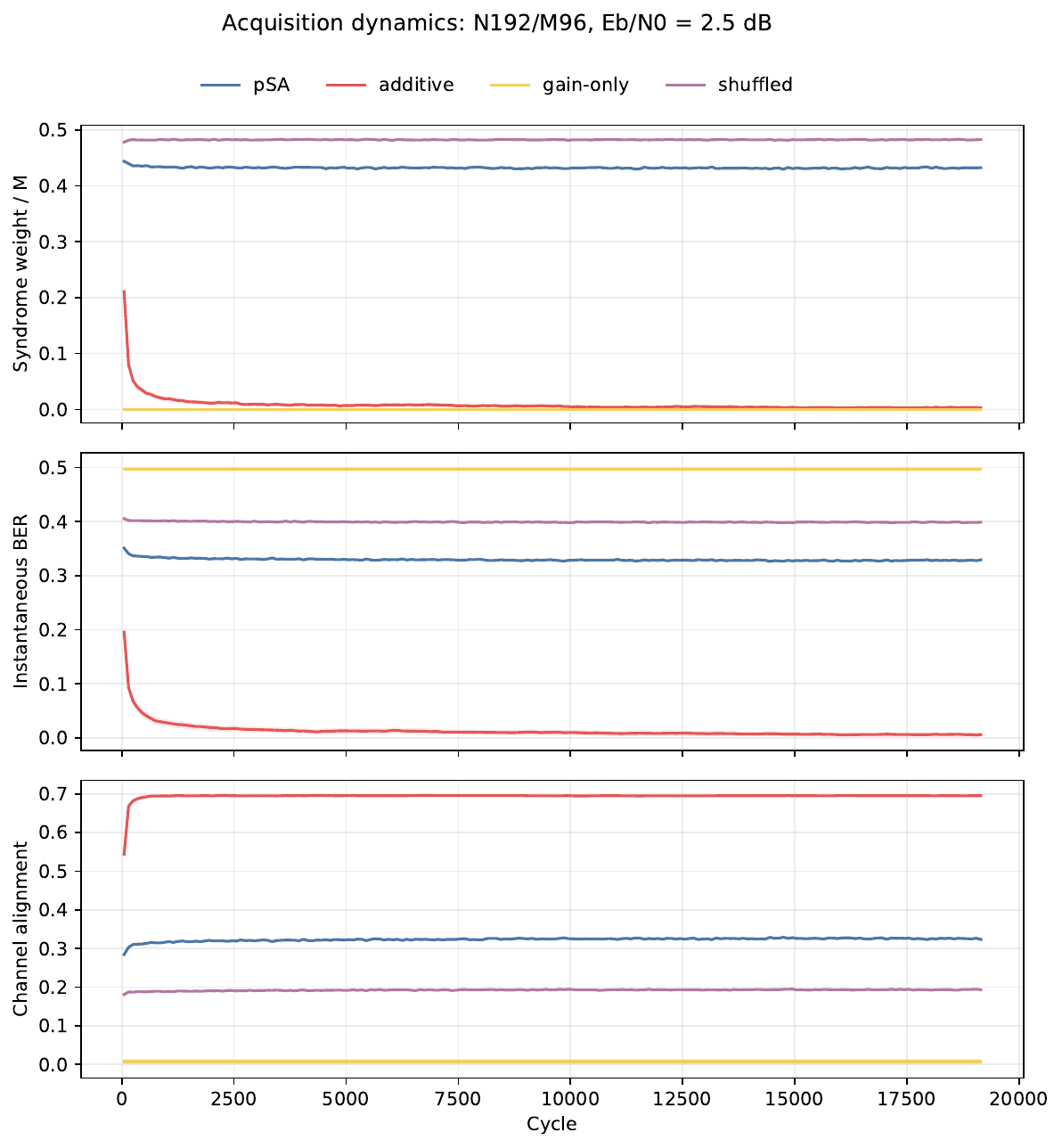}\\[1mm]
\includegraphics[width=0.81\textwidth]{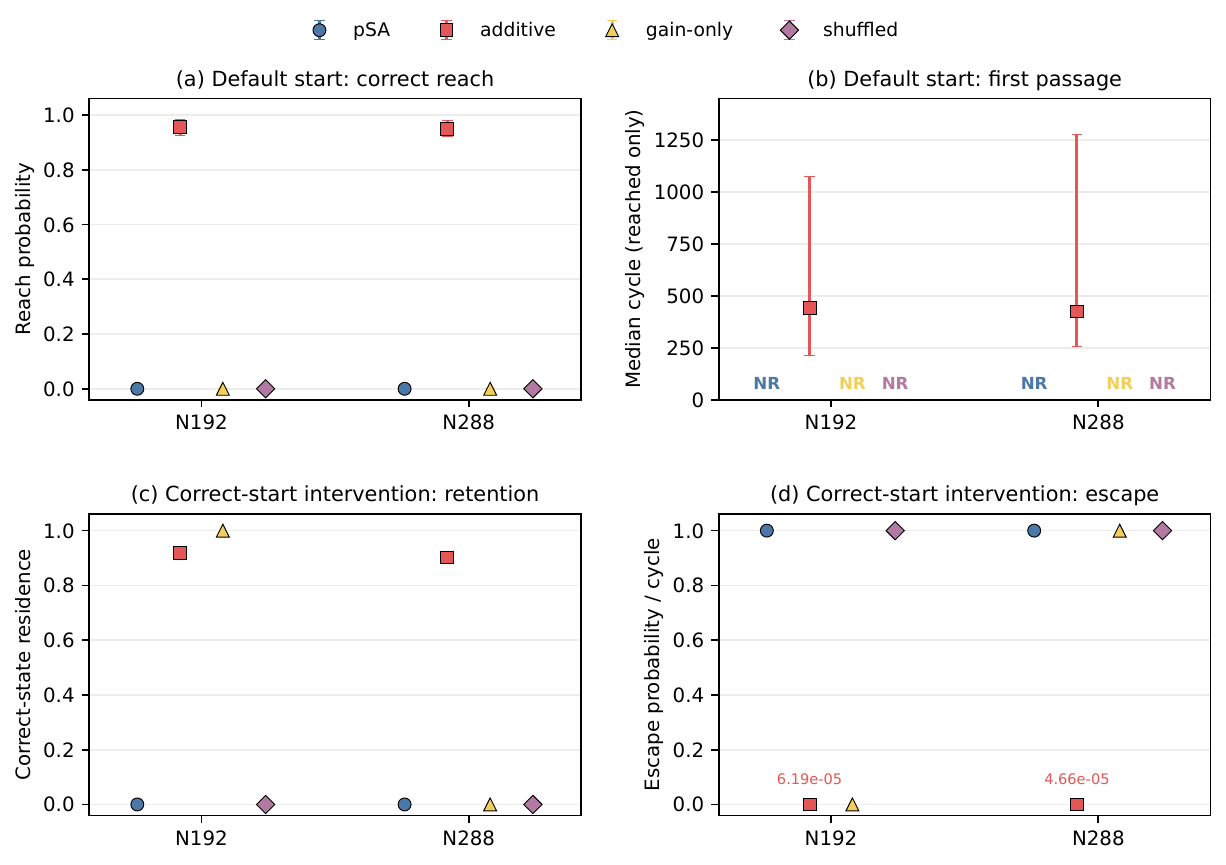}
\caption{Integrated mechanism evidence at $E_b/N_0=2.5$ dB on the representative matrices.  Top: default all-zero-initialized trajectories at $N=192$ jointly track syndrome, instantaneous BER, and channel alignment; gain-only can reach zero syndrome while remaining near BER 0.5.  Lower four panels: default-start reach and first-passage time (upper row) are separated from correct-start residence and escape probability (lower row); each intervention uses 200 trajectories per method and size.  Zero markers are true zero reach probabilities.  NR means not reached in any trajectory, so first-passage time is undefined.  Correct-start measures are retention interventions, not acquisition rates.  Alternative-initialization validation is reported in the Supplemental Material.}
\label{fig:mechanism}
\end{figure*}

\subsection{Post-acquisition stability and a wrong-codeword counterexample}

After an additive trajectory from the default initialization first reaches the correct word, its mean subsequent residence fraction is 0.99998 at both $N=192$ and 288.  A separate correct-start intervention initializes every method at the transmitted word with zero response memory, preventing acquisition selection from contaminating retention.  Under this intervention additive residence is 0.9176 and 0.9020.  Matched \psa{} and shuffled memory escape immediately and have zero residence at the sampled resolution.  Conditional on an additive escape, the return probabilities are approximately 0.959 and 0.963.

The microscopic decomposition is size dependent.  At $N=192$ and 288, reduced escape and rapid return both contribute.  At $N=96$, additive dynamics can escape more often than a comparator yet return much faster; across new $N=96$ codes, additive return times span roughly 49--84 cycles compared with 170--349 cycles for the matched \psa{} ablation.  We therefore use the general term \emph{post-acquisition stability}, not ``escape suppression,'' for the transferable mechanism.

Gain-only supplies a crucial counterexample.  At $N=192$ it can settle to zero syndrome with essentially zero late flips while maintaining BER near 0.5 and negligible alignment with the channel.  The score-selected binary-state control likewise reaches zero syndrome with negligible motion but acquires the transmitted word in none of 200 trajectories and has decoded BER 0.497.  Both have frozen in valid but wrong codewords.  At $N=288$, however, binary feedback at $\kappa=0.75$ acquires the transmitted word in 93\% of trajectories and retains it thereafter, while its decoded BER 0.01054 remains above the additive value 0.00595.  Generic persistence can therefore contribute, but low syndrome, low activity, or binary inertia alone does not explain the full channel-consistent decoding benefit.

\section{Robustness across independent LDPC code realizations}
\label{sec:robustness}

To test robustness across code realizations, one fixed size-specific additive package is transferred to ten independently generated random-regular matrices at each block length.  The same C00--C09 matrices, SNRs, stochastic seed design, channel generation, and initialization are used for every arm, with 1000 trials per code, SNR, and arm.  No package is retuned for an individual code.

For a separately optimized, package-level fixed-transfer comparator, we also transfer the size-specific \psa{} package independently selected on the representative matrix in Sec.~\ref{sec:performance}.  Each transferred package retains its complete selected dictionary, including schedule, activation, channel scaling, and readout.  Independently selected packages can therefore differ in readout and other nonmemory parameters; at $N=192$, for example, the additive package retains best-state readout whereas the pSA-specific package retains majority readout.  This comparison is interpreted at the package level.  The matched ablation instead holds the additive package's nonmemory parameters and readout fixed and changes only $\lambda$ to zero, thereby providing the rule-isolating comparison.

\begin{figure*}[t]
\centering
\includegraphics[width=0.98\textwidth]{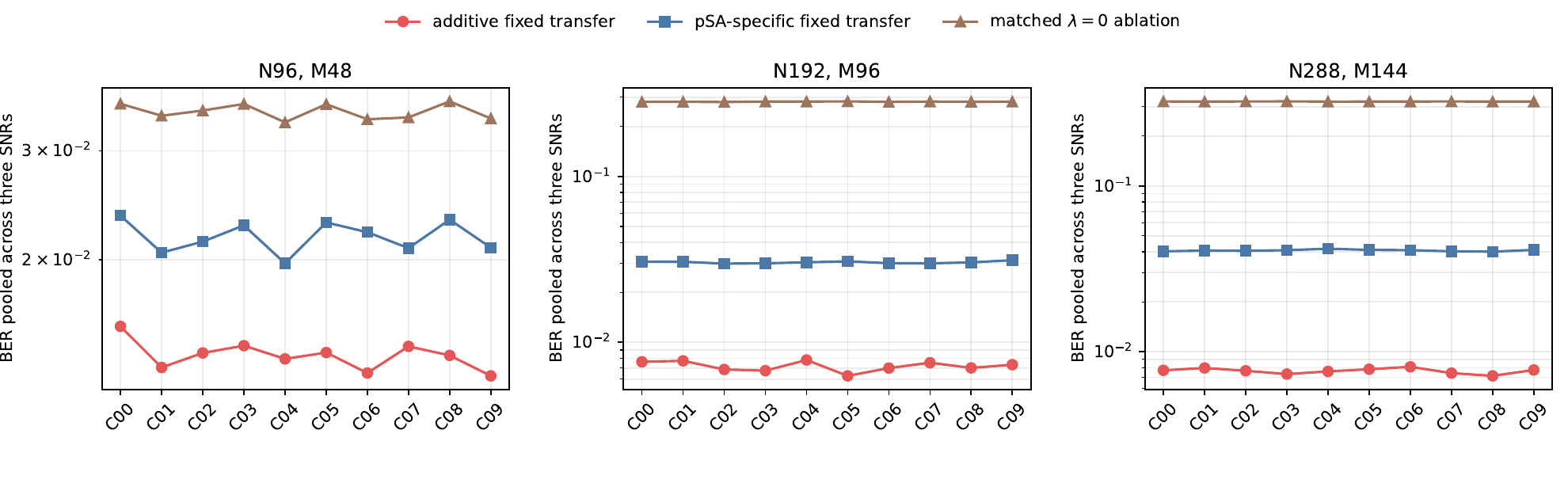}
\caption{Three-way fixed-transfer comparison across independent code realizations.  Each point is the BER for one code, pooled equally over 2.0, 2.5, and 3.0 dB.  The additive and pSA-specific packages were independently selected on the representative matrix and then transferred intact without per-code retuning; the matched $\lambda=0$ arm instead uses the additive nonmemory parameters and readout and removes only response reinforcement.  Ten codes are evaluated at each $N$ with 1000 trials per code, SNR, and arm.  The additive package has lower BER than both comparators for every code.  The pSA-specific arm is the package-level fixed-transfer performance baseline, whereas the matched ablation isolates the response-rule effect.  These data are not interpreted as finite-size scaling.}
\label{fig:crosscode}
\end{figure*}

Against the pSA-specific fixed-transfer package, the additive fixed-transfer package yields lower BER and FER for all 30 of 30 code realizations when pooled across the three SNRs (Fig.~\ref{fig:crosscode}).  The code-level median relative BER reductions and 95\% code-bootstrap intervals are 35.46\% $[32.29,38.83]\%$ at $N=96$, 76.61\% $[74.81,77.17]\%$ at $N=192$, and 81.10\% $[80.66,81.84]\%$ at $N=288$.  The independent code realization is the bootstrap unit.

The matched $\lambda=0$ ablation yields larger median BER reductions of 59.90\%, 97.46\%, and 97.61\%, respectively.  Those effects isolate the response-rule change under additive-optimized nonmemory parameters; they are not relabeled as independently optimized or pSA-specific performance margins.

Mechanism sampling on nine of these matrices (three per size) is consistent with the representative-code interpretation: additive dynamics have higher correct-basin acquisition and lower late correct-to-incorrect back-flip rate in all nine comparisons.  Detailed per-code tables, absolute error rates, matrix metadata, and trajectory summaries are in the Supplemental Material \cite{supplement}.

\section{Discussion}
\label{sec:discussion}

The controls identify saturated-response reinforcement as a rule-level resource rather than a scalar coefficient.  Additive memory changes both amplitude and history; neither ingredient alone suffices.  Destroying bit identity while preserving the delayed-response distribution removes the useful transition kernel.  The tuned binary-state control sharpens the conclusion: same-bit self-feedback can be beneficial, but it does not reproduce the full effect of retaining the real-valued saturated response.

For the tested LDPC decoders, the mechanism has two stages.  First, additive dynamics are substantially more likely to acquire the valid codeword consistent with channel evidence.  Second, after acquisition they remain near that basin or return rapidly after departure.  This distinction prevents a common mechanistic error.  Zero syndrome and low activity can also characterize a valid wrong-codeword freeze, as the $N=192$ gain-only and binary-state controls demonstrate.  The $N=288$ binary result additionally shows that persistence can materially assist acquisition without matching additive decoding quality.  A successful explanation must therefore include parity feasibility, channel consistency, and the retained variable.

Because the LDPC implementation resamples stochastic channel evidence at every cycle, part of the observed benefit may also reflect temporal integration of channel information in addition to stabilization of the parity-constrained parallel dynamics.  These contributions were not separately isolated; their separation under a deterministic channel field remains an open question.

This interpretation is consistent with, but distinct from, prior temporal-state approaches.  PIMI and time-dimensional exchange coupling reuse a bit's binary state \cite{zhu2026pimi,du2026tec}; spike-frequency adaptation constructs an adaptive real state from output activity and returns negative feedback on the local-field side \cite{xu2025sfa}; momentum methods retain configuration or continuous dynamical history \cite{okuyama2019momentum,brown2024momentumcim}.  The present rule retains the saturated response itself and adds it at the stochastic response path.  The contribution is therefore not memory or same-bit feedback per se, but evidence that the identity of the stored quantity and its reinjection rule determine computational function.

The 30-code experiment supports transfer across independently generated random-regular $(3,6)$ matrices under fixed size-specific packages.  Transferring the separately optimized pSA-specific package to the same matrices shows that the fixed-transfer advantage of the additive package is not merely a consequence of evaluating \psa{} under additive-optimized nonmemory parameters: the additive package wins in BER and FER for all 30 codes.  The matched $\lambda=0$ results remain a separate causal ablation.  The two comparisons answer different questions: package-level fixed-transfer performance and response-rule causality, respectively.  These studies do not establish universality over degree distributions, channels, schedules, or block lengths, and they are not finite-size scaling.  Likewise, initialization robustness was tested only on the representative $N=192$ and 288 matrices at 2.5 dB: random and channel-hard-decision starts preserve the acquisition advantage, while broader initial-state, channel, and ensemble dependence remains outside the present study.

From a physical-design perspective, finite p-bit settling and programmable coupling dynamics make retained circuit state a realistic co-design variable \cite{mcgoldrick2022settling,gibeault2024coupling}.  Equation~(\ref{eq:additive}) can be interpreted abstractly as a stored response plus an additive readout path, whereas Eq.~(\ref{eq:filter}) resembles an ideal first-order response node.  The data show that the response-path placement and identity of retained state can alter the transition kernel even when both algorithms add one value per bit.  They do not demonstrate a particular circuit realization, energy, area, or speed benefit, nor identify $\lambda$ or $\rho$ with a unique device parameter.  Precision, saturation, timing, interconnect delay, and variability remain implementation dependent.

Cross-problem boundary controls in the Supplemental Material do not support a generic advantage of temporal state under the tested conditions: the MAX-CUT results are co-optimized outcomes, whereas the random 2-SAT study is a matched-$k_w$ finite-response-$\rho$ control.  These studies delimit generalization rather than provide additional applications or a matched causal test of the additive LDPC rule \cite{supplement}.

\raggedbottom
\section{Conclusion}
\label{sec:conclusion}

Bit-aligned additive saturated-response memory creates a distinct temporal-reinforcement regime in parallel p-bit LDPC dynamics.  It improves independently optimized representative-code BER and FER, survives gain, normalization, shuffling, and tuned binary-state controls, promotes channel-consistent correct-basin acquisition from all-zero, random, and channel-hard-decision starts, and improves post-acquisition stability.  Fixed additive packages also outperform separately optimized pSA-specific packages transferred without per-code retuning on all 30 independent code realizations; the larger matched $\lambda=0$ effects separately support response-rule causality.  Binary self-feedback can contribute but does not reproduce the saturated-response benefit.  These results motivate co-design of the retained variable and its response-path reinjection around computational structure.

\section*{Data Availability Statement}

The simulation code, processed data, parameter files, and metadata supporting this study are publicly available in the GitHub repository and Zenodo archive cited in Ref.~\cite{onizawa_psa_ldpc_zenodo_2026}.  The records include the binary-state-control, three-arm fixed-transfer, and initialization-robustness validations and the scripts for reproducing the figures from processed data.

\section*{Acknowledgments}

OpenAI ChatGPT (GPT-5.6 Sol) and OpenAI Codex were also used for literature organization and drafting and revision of the manuscript.  N.~Onizawa directed, reviewed, and refined all AI-assisted material, checked the cited references, and takes full responsibility for the content.  Tool identifiers and research-related assistance are disclosed in Appendix~\ref{app:parameters}.  No AI system is listed as an author.

\onecolumngrid
\clearpage
\twocolumngrid
\flushbottom
\appendix

\section{LDPC sign convention and stochastic channel field}
\label{app:mapping}

The simulator stores binary bits $b_i\in\{0,1\}$ and thresholds them to one when the signed decision variable is nonnegative.  The corresponding manuscript spin is therefore $x_i=2b_i-1$.  For a degree-$d_a$ check,
\begin{equation}
\prod_{j\in\partial a}x_j=(-1)^{d_a-\sum_{j\in\partial a}b_j}.
\label{eq:checkspin}
\end{equation}
Because the tested check degree is $d_a=6$, even binary parity is equivalent to $\prod_jx_j=+1$.  Holding $x_i$ out of the product gives the first term of Eq.~(\ref{eq:ldpcfield}); in the binary kernel this same quantity is evaluated as the XOR of the other five bits and mapped to $+1$ for XOR one and $-1$ for XOR zero.

For channel value $z_i$, the comparison $r_i=\mathbf 1[z_i<u_i]$ with $u_i\sim\mathcal U[-1,1]$ gives
\begin{equation}
\Pr(r_i=1)=\frac{1-z_i}{2},\qquad
\E[2r_i-1]=-z_i.
\label{eq:channelmean}
\end{equation}
For BPSK, a positive received sample supports binary zero, hence $z_i>0$ gives a negative expected field and favors $x_i=-1$ as required.  Equations~(\ref{eq:ldpcenergy}) and (\ref{eq:ldpcfield}) are therefore the exact cycle-conditioned objective and drive for the implementation.  Their channel expectation is obtained by replacing $c_i^{(t)}$ with $-z_i$.  The factorized form is preferable to calling the model a pairwise Hamiltonian because each degree-six parity check is a sixth-order product and the channel factor is resampled each cycle.

A decoded state is called (i) \emph{valid} if all check products equal $+1$, (ii) \emph{correct} if it also equals the transmitted word, and (iii) \emph{channel consistent} when its signed channel score $\sum_i z_i(1-2b_i)=-\sum_i z_i x_i$ is high.  This separation is essential because validity alone does not select the transmitted codeword.

\section{Parameter provenance and readout}
\label{app:parameters}

OpenAI ChatGPT (GPT-5.6 Sol) and OpenAI Codex (recorded model identifiers \texttt{gpt-5.6-sol} and \texttt{gpt-6-astra}; Desktop client version 26.901.51231 at final preparation) assisted code development and debugging and the development and review of analysis procedures.  N.~Onizawa directed this work, executed and validated all simulations and analyses, and checked numerical claims against the archived outputs.

Table~\ref{tab:transferparams} records the key fields of the additive packages used for cross-code transfer.  Every value except $\lambda$ is shared by the additive and matched-ablation arms at a given size.  The pSA-specific transfer instead uses the complete parameter-and-readout package independently selected for \psa{} on the representative matrix; its side-by-side values are given in the Supplemental Material.  The matched causal controls and acquisition--stability analysis use the size-specific additive-optimized parameter sets on the representative matrices, with the rule changes described in Secs.~\ref{sec:controls} and \ref{sec:mechanism}.  Complete machine-readable dictionaries, seeds, configuration hashes, and reproduction scripts are available in the public repository and archive \cite{onizawa_psa_ldpc_zenodo_2026}.

\begin{table}[t]
\caption{Key fields of the fixed additive packages used for cross-code transfer.  The matched \psa{} ablation uses $\lambda=0$ with all remaining entries, including readout, unchanged.}
\label{tab:transferparams}
\begin{ruledtabular}
\begin{tabular}{lccc}
Field & $N=96$ & $N=192$ & $N=288$ \\
\hline
$\lambda$ & 0.95 & 0.95 & 0.90 \\
cycles & 25,600 & 19,200 & 25,600 \\
schedule & piecewise & cosine & cosine \\
$I_{0,\min}$ & 0.06227 & 0.07745 & 0.22683 \\
$I_{0,\max}$ & 1.20690 & 0.73899 & 8.36851 \\
$p_{\mathrm{hold}}$ & 0.73137 & 0.46613 & 0.40758 \\
$k_w$ & 4.61543 & 3.84760 & 0.17046 \\
readout & majority & best state & best state \\
\end{tabular}
\end{ruledtabular}
\end{table}

Response state is initialized to zero unless an intervention explicitly states otherwise.  The benchmark, matched-control, and default-initialization mechanism kernels initialize every stored LDPC bit to zero; this is a valid codeword but is generally not the randomly transmitted word.  The initialization-robustness intervention instead uses either independent equiprobable bits or the hard decision from the received channel sample while maintaining paired initial states across methods.  The correct-start retention challenge initializes bits at the transmitted word while retaining zero response memory.  These interventions answer acquisition robustness and conditional stability questions and are never pooled.

The available readouts are the final state, a bitwise majority over the selected post-burn-in window, and a best-visited-state rule.  For LDPC, best state first minimizes syndrome weight and then maximizes $\sum_i z_i(1-2b_i)$ among ties.  For MAX-CUT it maximizes $C(\bm x)$, and for 2-SAT it minimizes the unsatisfied-clause count.  The selected rule, burn-in, and sampling window are part of each recorded parameter set.

\section{Statistical definitions}
\label{app:statistics}

For $T$ trials of length $N$, $\mathrm{BER}=\sum_{u=1}^{T}d_H(\hat{\bm b}_u,\bm b_u)/(TN)$ and $\mathrm{FER}=T^{-1}\sum_u\mathbf{1}[\hat{\bm b}_u\neq\bm b_u]$.  Relative reduction is $1-y_{\mathrm{add}}/y_{\mathrm{ctrl}}$.  In the matched causal-control and initialization-robustness studies, common seed streams define paired batch differences, and intervals are computed across the ten prespecified seed batches.  The initialization acquisition-difference intervals use the same unbounded paired-$t$ construction and are not clipped to the probability range.  In the cross-code transfer study, code-level pooled ratios are formed before resampling; the independent code realization, not the trial or SNR point, is the bootstrap unit.  Percentile intervals use 20,000 bootstrap resamples.

For the matched-$k_w$ 2-SAT control, each formula contributes the mean outcome over its 100 paired trials at a fixed $(k_w,\rho)$ cell.  Formula-wise treatment-minus-control differences are resampled with replacement to form the reported 95\% percentile interval.  Satisfiable and unsatisfiable strata are analyzed separately only as secondary outcomes.

\section{Acquisition and stability metrics}
\label{app:metrics}

Let $T_{\mathrm{hit}}=\inf\{t:\bm{x}^{(t)}=\bm{x}^{\star}\}$ be the first cycle at which the transmitted codeword $\bm{x}^{\star}$ is visited.  Correct-reach probability is the fraction of trajectories with finite $T_{\mathrm{hit}}$ inside the simulation horizon.  For a reaching trajectory, post-hit residence is
\begin{equation}
R=\frac{1}{T-T_{\mathrm{hit}}+1}\sum_{t=T_{\mathrm{hit}}}^{T}
\mathbf{1}[\bm{x}^{(t)}=\bm{x}^{\star}].
\end{equation}
An escape is a transition from $\bm{x}^{\star}$ to any other state; return time counts cycles until the next visit.  A back-flip is a correct-to-incorrect transition relative to the transmitted word, not an arbitrary bit flip.  For each trajectory, its rate is the number of such transitions divided by the number of previously correct bit states in the analysis window, including held bits.  Counts are pooled within the window before division, not averaged as cycle-wise ratios; zero denominators yield an undefined value omitted from finite-value summaries.  Late rates use the final quarter of cycles (with the recorded bin boundary) and are then averaged over trajectories.  Total bit-flip rate instead counts all reversals per bit-cycle; the separately logged back-flip probability counts only correct-to-incorrect transitions per bit-cycle.  Supplemental Sec.~S5 gives the binning and aggregation details.  ``Post-acquisition stability'' refers jointly to residence, escape, and return behavior and does not assume that the same microscopic component dominates at every $N$.

\bibliography{references}

\end{document}


\title{Supplemental Material for ``Error-Rate Reduction in LDPC Decoding via Bit-Aligned Temporal Reinforcement in Parallel Probabilistic-Bit Dynamics''}
\author{Naoya Onizawa}
\affiliation{Research Institute of Electrical Communication, Tohoku University, Sendai 980--8577, Japan}
\author{Takahiro Hanyu}
\affiliation{Research Institute of Electrical Communication, Tohoku University, Sendai 980--8577, Japan}
\date{September 8, 2026}
\maketitle

This Supplemental Material reports the full control, trajectory, transfer, cross-problem, and validation evidence underlying the LDPC-centered main manuscript.  The independently optimized representative-code benchmark, matched-parameter interventions, binary-state control with only $\kappa$ separately swept, pSA-specific fixed transfer, matched $\lambda=0$ ablation, initialization-robustness validation, and matched-$k_w$ finite-response 2-SAT control remain separate throughout.

\section{Study design and parameter audit}

\begin{table}[ht]
\caption{Study-level reproducibility records and their scientific roles.}
\label{tabS:provenance}
\small
\begin{tabular}{@{}p{0.18\textwidth}p{0.34\textwidth}p{0.38\textwidth}@{}}
\toprule
\raggedright Study & \raggedright Archived machine-readable information & \raggedright Design role \tabularnewline
\midrule
\raggedright Independently optimized LDPC benchmark & \raggedright Per-mode outcomes, parameter sets, random seeds, and representative-matrix identifiers & \raggedright Best-achieved high-statistics performance on one fixed matrix per block length \tabularnewline
\raggedright Matched causal controls & \raggedright Condition-level outcomes, paired seed-batch statistics, and coefficient sweeps & \raggedright Rule discrimination with the additive-optimized nonmemory parameters held fixed \tabularnewline
\raggedright Binary-state self-feedback & \raggedright $\kappa$ sweep, equal-range endpoints, high-statistics selected comparators, paired statistics, and trajectory summaries & \raggedright Test whether same-bit binary inertia is sufficient to reproduce saturated-response reinforcement \tabularnewline
\raggedright Response-alignment and trajectory diagnostics & \raggedright Condition, seed-batch, and trajectory summaries; derangement checks & \raggedright Bit-specific response alignment and dynamical observables \tabularnewline
\raggedright Acquisition and stability & \raggedright All-zero, random, channel-hard, and correct-start trajectory summaries; paired initialization metadata & \raggedright Correct-basin acquisition, initialization robustness, and conditional post-acquisition stability \tabularnewline
\raggedright Cross-code transfer robustness & \raggedright Three-arm code-level outcomes, complete packages, seeds, matrix hashes, and rank checks & \raggedright Package-level pSA-specific fixed-transfer comparison and separate matched $\lambda=0$ rule ablation \tabularnewline
\raggedright Matched-$k_w$ 2-SAT control & \raggedright Formula-level response surface, paired statistics, formula identifiers, and satisfiability labels & \raggedright Independent finite-response effect after removing $k_w$ coupling \tabularnewline
\bottomrule
\end{tabular}
\end{table}

For the cross-code transfer study, complete size-specific parameter dictionaries and seed records are publicly available in machine-readable form \cite{onizawa_psa_ldpc_zenodo_2026}.  Table~\ref{tabS:parameters} compares the additive and pSA-specific packages; values shown to more digits are recorded parameter values, not claims of physical precision.  The matched $\lambda=0$ ablation uses the additive column at each size with only $\lambda$ set to zero.

\clearpage

\begin{table}[ht]
\caption{Fixed additive and pSA-specific packages used for cross-code transfer.  Additive (A) and pSA-specific (P) packages were independently selected on the representative matrix and transferred without per-code retuning.  The matched ablation uses A with $\lambda=0$.}
\label{tabS:parameters}
\scriptsize
\begin{ruledtabular}
\begin{tabular}{lcccccc}
& \multicolumn{2}{c}{$N=96$} & \multicolumn{2}{c}{$N=192$} & \multicolumn{2}{c}{$N=288$} \\
Field & A & P & A & P & A & P \\
\hline
$\lambda$ & 0.95 & 0 & 0.95 & 0 & 0.90 & 0 \\
cycles & 25,600 & 25,600 & 19,200 & 19,200 & 25,600 & 25,600 \\
schedule & piecewise & exponential & cosine & linear & cosine & linear \\
$I_{0,\min}$ & 0.062274 & 0.030036 & 0.077447 & 0.329820 & 0.226829 & 0.288425 \\
$I_{0,\max}$ & 1.20690 & 8.90735 & 0.738992 & 7.86812 & 8.36851 & 9.23865 \\
schedule shape & 0.616804 & 1.14992 & 0.124214 & 0.050627 & 0.069103 & 4.33860 \\
initial plateau & 0.171044 & 0.247503 & 0.564513 & 0.233693 & 0.383346 & 0.826974 \\
$p_{\mathrm{hold}}$ & 0.731367 & 0.669070 & 0.466129 & 0.786535 & 0.407577 & 0.843957 \\
$k_w$ & 4.61543 & 6.07160 & 3.84760 & 0.343024 & 0.170457 & 6.57492 \\
$k_r$ & 4.95663 & 8.70124 & 3.92867 & 0.326247 & 0.169291 & 6.67830 \\
channel scaling & LLR: 0.5380 & SNR: 1.4080 & SNR: 0.8017 & LLR: 1.0318 & fixed: 1.7204 & LLR: 0.9567 \\
burn-in & 3,204 & 13,349 & 13,837 & 12,836 & 25,389 & 10,774 \\
sample window & 11,294 & 3,503 & 5,306 & 3,856 & 196 & 7,843 \\
readout & majority & majority & best state & majority & best state & best state \\
\end{tabular}
\end{ruledtabular}
\end{table}

In the source records, \texttt{I0\_hold\_fraction} denotes the initial-schedule plateau fraction and affects $I_0^{(t)}$ only for the piecewise schedule; the stored cosine-schedule values are inactive.  By contrast, \texttt{psa\_p} denotes $p_{\mathrm{hold}}$, the independent per-bit, per-cycle probability of suppressing both state and response-state commitment, with activation probability $1-p_{\mathrm{hold}}$.  These quantities are not interchangeable.

The ten stochastic seeds used in the cross-code study are generated as $20260826+8022j$ for $j=0,\ldots,9$.  The same code matrices, channel and stochastic seed design, and initial states are used for all three arms.  Each independently selected package retains its own recorded readout; in particular, the $N=192$ pSA-specific package uses majority readout while the additive and matched-ablation packages use best-state readout.  Matrix-generation seeds and SHA-256 hashes of every matrix are retained in the code-level metadata.  The additive/matched configuration record has hash prefix \texttt{f96f4a3806b4}.

\section{Independently optimized representative LDPC benchmark}

\begin{figure}[ht]
\centering
\includegraphics[width=0.72\textwidth]{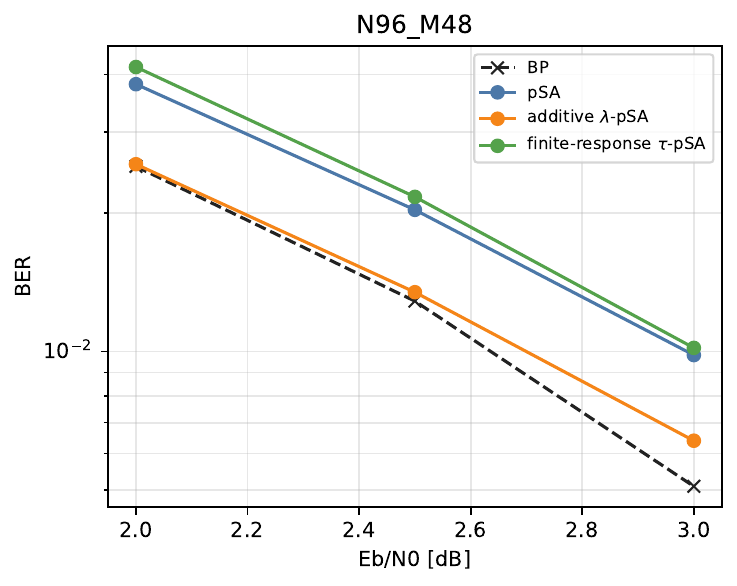}
\caption{Representative $N=96$ BER.  The three p-bit modes are independently optimized and use 10,000 trials per SNR; their selection and evaluation protocol is specified in Sec.~S2A.  The conventional BP reference uses 5000 trials per SNR from a separate Monte Carlo stream, is not paired trial by trial to the p-bit modes, and does not undergo their parameter search.}
\label{figS:N96}
\end{figure}

The representative-code benchmark independently optimizes \psa{}, additive $\lambda$-pSA, and finite-response $\tau$-pSA for one fixed matrix at each size.  Figure~\ref{figS:N96} supplies the $N=96$ BER panel omitted from the five-figure main-text layout.  Its comparison design is the same as main Fig.~2, not the matched-parameter intervention.

\subsection{Representative-code parameter optimization}

For every block length and algorithmic mode, a separate Optuna tree-structured Parzen estimator search evaluated 1000 candidate packages at $E_b/N_0=2.0$, 2.5, and 3.0 dB.  Each initial candidate used 160 trials per SNR and one candidate-specific stochastic stream.  The search varied $k_w$, $k_r$, channel scaling and its fixed/SNR/LLR convention, $p_{\mathrm{hold}}$, cycle count, $I_{0,\min}$, $I_{0,\max}$, schedule family and shape, initial plateau fraction, readout, burn-in, and sampling window; it additionally varied $\lambda\in[0,0.95]$ for additive dynamics or $\rho\in[0,0.99]$ for finite-response dynamics.  The continuous domains were $k_w\in[0.05,8]$, $k_r\in[0.05,12]$, channel-scale coefficient $\alpha\in[0.05,8]$, $I_{0,\min}\in[0.03,3]$, $I_{0,\max}\in[I_{0,\min},12]$ (all log sampled), schedule shape in $[0.05,10]$ (log sampled), initial plateau fraction in $[0,0.9]$, and $p_{\mathrm{hold}}\in[0,1]$ for \psa{} and additive $\lambda$-pSA or $[0,0.8]$ for finite-response $\tau$-pSA.  Schedule choices were linear, constant, exponential, cosine, and piecewise; readout choices were final state, post-burn-in majority, and best visited state.  The cycle choices were $\{9600,12800,19200,25600\}$ for $N=96$ and 288 and $\{6400,9600,12800,19200\}$ for $N=192$.

Candidates were ranked by
\begin{align*}
S={}&\left\langle\left|\log_{10}(B_e+\epsilon)-\log_{10}(B_{\mathrm{BP},e}+\epsilon)\right|\right\rangle_e \notag\\
&+0.2\left\langle\left|\log_{10}(F_e+\epsilon)-\log_{10}(F_{\mathrm{BP},e}+\epsilon)\right|\right\rangle_e
+0.5\left\langle V_e\right\rangle_e,
\end{align*}
where $B_e$, $F_e$, and $V_e$ are BER, FER, and the nonzero-syndrome fraction at SNR index $e$, $\epsilon=10^{-12}$, and the brackets denote an equal average over the three SNRs.  Ties in $S$ were resolved by the BER log-gap term; no additional scientific tie-break was specified for an exact residual tie.  The 32 lowest-score candidates were each reevaluated with 2000 total trials per SNR divided across six new seed streams.  The lowest-score refined package was then evaluated with 10,000 total trials per SNR divided across ten further seed streams.

For \psa{}, this ten-stream evaluation is the reported high-statistics endpoint.  For the two response-state modes, the response coefficient alone was subsequently scanned while all other selected fields were fixed: a coarse grid from 0 to 0.95 in steps of 0.05 was combined with a step-0.02 grid within $\pm0.15$ of the confirmed coefficient, subject to the mode bounds.  Every grid point used 10,000 trials per SNR and its own sweep stream; the minimum-$S$ point, with the same BER-gap tie break and then ascending coefficient order for an exact residual tie, supplies the reported additive or finite-response endpoint.  Thus search, refinement, and coefficient-sweep streams are distinct, but the displayed response-state endpoint is the selected coefficient-grid evaluation itself rather than an additional held-out run.  Complete candidate-level search and seed records are available in the public release \cite{onizawa_psa_ldpc_zenodo_2026}.

\subsection{BP reference decoder}

The BP reference is a custom NumPy implementation of the binary sum-product algorithm on the same fixed representative parity-check matrix used by the p-bit benchmark.  It uses a flooding schedule: variable-to-check messages are initialized to the clipped channel LLR, check-to-variable messages to zero, and every check update is completed before the variable and posterior updates of that iteration.  Channel and edge messages are clipped to $[-50,50]$, and the product entering $\operatorname{atanh}$ is clipped to $[-1+10^{-12},1-10^{-12}]$.  A posterior LLR below zero is decoded as bit one; a zero syndrome after a complete flooding iteration terminates decoding early, otherwise the decoder stops after 50 iterations.  With BPSK $0\mapsto+1$, $1\mapsto-1$ and code rate $R=\dim(G)/N$, the AWGN variance is $\sigma^2=[2R10^{(E_b/N_0)/10}]^{-1}$ and the channel LLR is $L_i=2y_i/\sigma^2$.  The BP curve uses 5000 random messages and noise realizations per SNR.  It shares the matrix, rate definition, channel model, and SNR values with the p-bit benchmark but uses separate Monte Carlo streams and is not paired trial by trial; conditional on the received vector, the decoder itself has no stochastic update.  No external decoding library is used.

\begin{table}[ht]
\caption{Exact independently optimized representative-code outcomes by SNR.}
\label{tabS:benchmark}
\begin{ruledtabular}
\begin{tabular}{cccccc}
$N$ & $E_b/N_0$ & \multicolumn{2}{c}{BER} & \multicolumn{2}{c}{FER} \\
 & (dB) & \psa{} & additive & \psa{} & additive \\
\hline
96 & 2.0 & 0.0381167 & 0.0255354 & 0.3856 & 0.2617 \\
96 & 2.5 & 0.0203219 & 0.0134563 & 0.2466 & 0.1772 \\
96 & 3.0 & 0.00982292 & 0.00639792 & 0.1525 & 0.1166 \\
192 & 2.0 & 0.0546740 & 0.0176484 & 0.5461 & 0.1314 \\
192 & 2.5 & 0.0264776 & 0.00419792 & 0.2888 & 0.0375 \\
192 & 3.0 & 0.00958646 & 0.00100990 & 0.1145 & 0.0111 \\
288 & 2.0 & 0.0770594 & 0.0178858 & 0.7020 & 0.1535 \\
288 & 2.5 & 0.0342743 & 0.00347292 & 0.3805 & 0.0346 \\
288 & 3.0 & 0.00941979 & 0.000564931 & 0.1303 & 0.0066 \\
\end{tabular}
\end{ruledtabular}
\end{table}

\section{Matched causal controls}

The matched causal-control study fixes the size-specific additive-optimized parameter set, shares seed streams across variants, and compares additive, matched \psa{}, normalized memory, and gain-only.  Table~\ref{tabS:controls} reports the 10,000-trial evaluation pooled equally over the three SNRs.  The independently optimized \psa{} column is included only to connect the matched-parameter evaluation with the representative-code benchmark; it is not paired to the control variants.

\begin{table}[ht]
\caption{High-statistics matched-control summary.}
\label{tabS:controls}
\begin{ruledtabular}
\begin{tabular}{lcccccc}
$N$ & $\lambda$ & Optimized \psa{} & Additive & Matched \psa{} & Normalized & Gain-only \\
\hline
96  & 0.95 & 0.022754 & 0.015403 & 0.035198 & 0.060372 & 0.031136 \\
192 & 0.95 & 0.030246 & 0.007365 & 0.281805 & 0.315129 & 0.499692 \\
288 & 0.90 & 0.040251 & 0.007597 & 0.322369 & 0.338275 & 0.322310 \\
\end{tabular}
\end{ruledtabular}
\end{table}

The paired control-minus-additive BER intervals over seed batches are positive in all nine comparisons.  The additive coefficient sweep shows an $N=96$ plateau over approximately 0.90--1.25, an $N=192$ optimum near 0.95, and an $N=288$ optimum near 0.90.  No size shows continuing improvement at the enlarged upper boundary.  At $\lambda=0$, all four kernels are trajectory-identical under common random streams.  Independent additive implementations agree trajectory by trajectory throughout the evaluated $\lambda\leq0.95$ range.

\section{Binary-state self-feedback control}

The stored-quantity control uses the same Jacobi pre-update bit state as the local-field evaluation,
\begin{equation}
d_{i,\mathrm{bin}}^{(t)}=q_i^{(t)}+\kappa x_i^{(t)}.
\label{eqS:binaryfeedback}
\end{equation}
There is no separate response state in this rule.  A held bit keeps its binary state under the common activation mask, and $\kappa=0$ is trajectory-identical to matched \psa{} under common random streams.  Equation~(\ref{eqS:binaryfeedback}) is a PIMI-inspired stored-quantity control, not a reproduction of PIMI: the noise distribution, schedule, stochastic partial activation, and LDPC workload differ.

\begin{figure}[ht]
\centering
\includegraphics[width=0.94\textwidth]{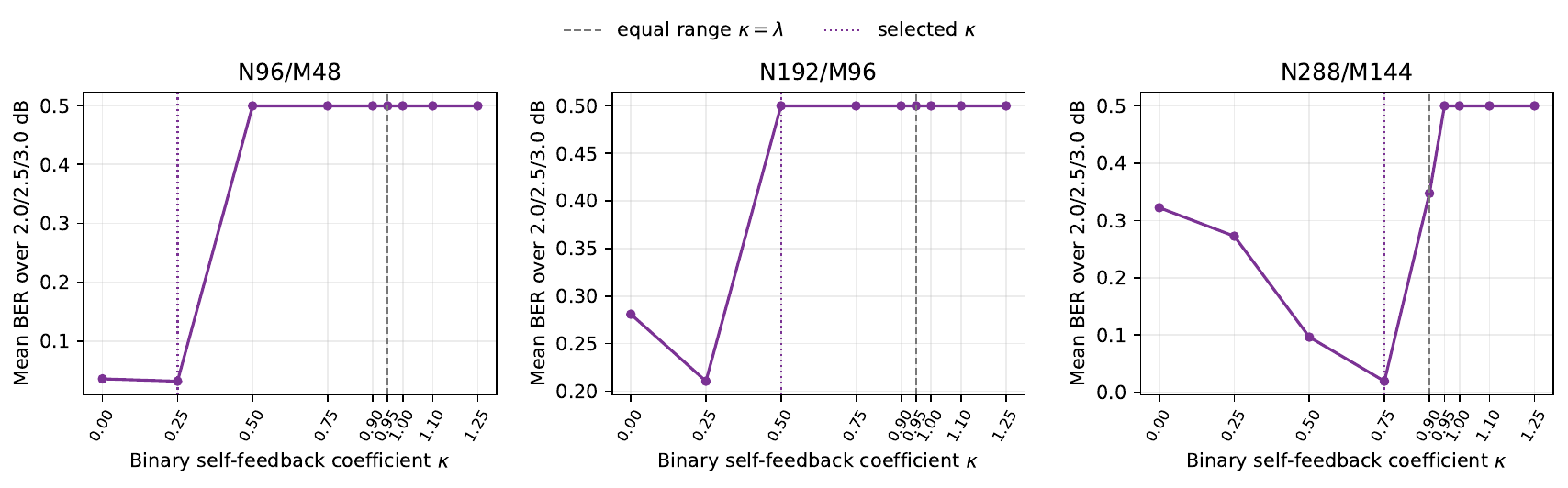}
\caption{Binary-state self-feedback BER as a function of $\kappa$ under the fixed matched-control nonmemory parameters.  The prespecified score selects $\kappa=0.25$, 0.50, and 0.75 for $N=96,192,288$, respectively.}
\label{figS:binary-sweep}
\end{figure}

\begin{figure}[ht]
\centering
\includegraphics[width=0.94\textwidth]{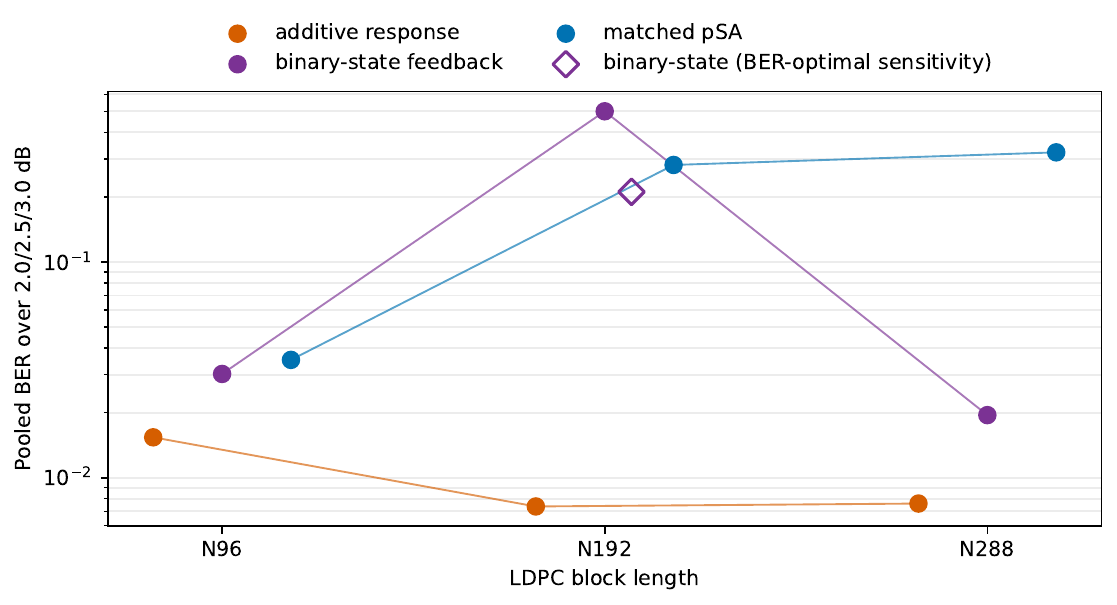}
\caption{High-statistics additive, binary-state, and matched-\psa{} comparison shown as points and lines.  The regular binary-state series uses the score-selected comparator: at $N=192$ its purple circular markers represent $\kappa=0.50$.  The additional open diamond markers represent the separately validated, BER-favorable $\kappa=0.25$ sensitivity within the tested coefficient sweep, not a global optimum or the original score selection.  Table~\ref{tabS:binary-best} uses that $N=192$ sensitivity for the publication-facing comparison; both outcomes are retained here.}
\label{figS:binary-highstat}
\end{figure}

Two estimands are retained separately.  The equal-range comparison sets $\kappa=\lambda$ so that binary and additive rules have the same maximum deterministic range.  The best-performance comparison sweeps $\kappa$ and then subjects the selected binary comparator to the same 10,000-trial-per-SNR evaluation used for publication-facing controls.  The prespecified sweep score was
\begin{align*}
S_\kappa={}&\left\langle\left|\log_{10}(B_{\kappa,e}+\epsilon)
-\log_{10}(B_{\mathrm{BP},e}+\epsilon)\right|\right\rangle_e \\
&+0.2\left\langle\left|\log_{10}(F_{\kappa,e}+\epsilon)
-\log_{10}(F_{\mathrm{BP},e}+\epsilon)\right|\right\rangle_e
+0.5\left\langle V_{\kappa,e}\right\rangle_e,
\end{align*}
where $B$, $F$, and $V$ denote BER, FER, and the nonzero-syndrome fraction, respectively, $\epsilon=10^{-12}$, and $\langle\cdot\rangle_e$ is an equal average over $E_b/N_0=2.0$, 2.5, and 3.0 dB after pooling the 2000 trials per SNR across seed batches.  The minimum over the predeclared $\kappa$ grid was selected; ties were broken by the BER-only log-gap term and then by grid order.  Figure~\ref{figS:binary-sweep} reports selection, and Fig.~\ref{figS:binary-highstat} reports the high-statistics endpoint comparison.

\begin{table}[ht]
\caption{Publication-facing binary-state self-feedback comparison pooled equally over the three SNRs.  Differences are binary minus additive BER with 95\% paired seed-batch intervals.}
\label{tabS:binary-best}
\begin{ruledtabular}
\begin{tabular}{ccccc}
$N$ & $\kappa$ & Binary BER & Additive BER & Difference [95\% interval] \\
\hline
96  & 0.25 & 0.0302868 & 0.0154035 & 0.0148833 $[0.0142088,0.0155578]$ \\
192 & 0.25 & 0.211519  & 0.0073646 & 0.204155 $[0.203056,0.205253]$ \\
288 & 0.75 & 0.0195231 & 0.0075966 & 0.0119265 $[0.0112603,0.0125927]$ \\
\end{tabular}
\end{ruledtabular}
\end{table}

At $N=96$ and 288, the score-selected values are also the BER- and FER-minimizing values on the evaluated grid.  At $N=192$, the score selects $\kappa=0.50$, but trajectory analysis shows a valid wrong-codeword freeze.  To avoid presenting that pathological score selection as the only binary comparator, the publication-facing table instead uses the separately high-statistics-validated, BER-favorable $\kappa=0.25$ sensitivity.  This substitution is conservative for the binary rule and is unique to $N=192$; it does not alter the archived score selection.

The equal-range settings are $\kappa=\lambda=0.95$ for $N=96$ and 192 and $\kappa=\lambda=0.90$ for $N=288$.  Their binary BERs are 0.49951, 0.49981, and 0.34739, compared with additive BERs 0.01576, 0.00749, and 0.00740 and matched-\psa{} BERs 0.03581, 0.28087, and 0.32236.  Thus amplitude matching does not reproduce the additive rule, but these endpoints are not substituted for the best-$\kappa$ estimand in Table~\ref{tabS:binary-best}.

\begin{figure}[ht]
\centering
\includegraphics[width=0.94\textwidth]{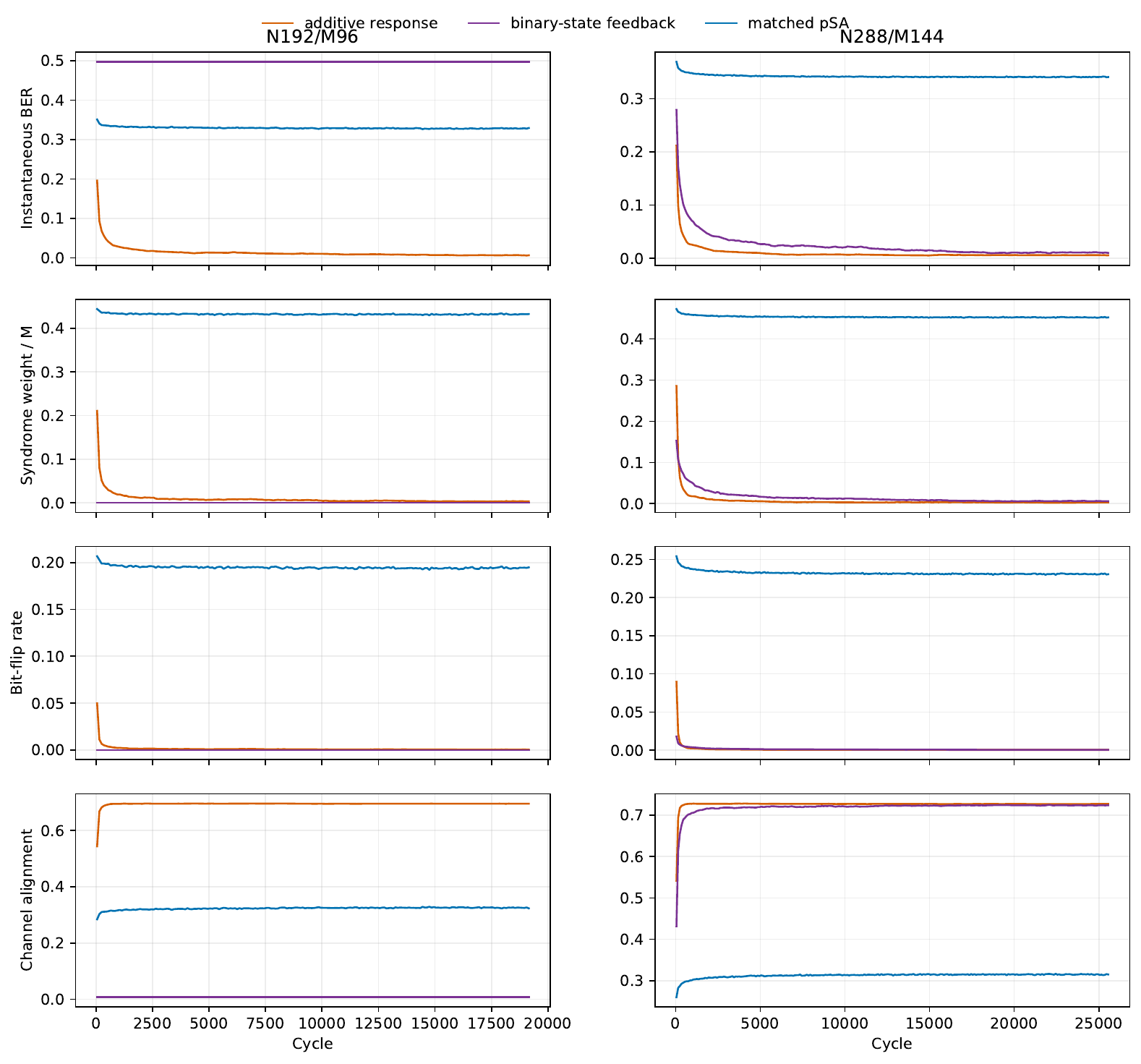}
\caption{Targeted binary-state trajectory evidence at 2.5 dB.  At $N=192$, score-selected $\kappa=0.50$ reaches a motionless valid but wrong codeword; at $N=288$, $\kappa=0.75$ materially improves acquisition but remains less accurate than additive saturated-response reinforcement.}
\label{figS:binary-trajectory}
\end{figure}

With 200 trajectories per method and size, the $N=192$ score-selected binary rule acquires the transmitted word in 0\% of trajectories, has decoded and late BER about 0.497, and reaches zero syndrome with negligible late motion.  At $N=288$, binary feedback acquires the transmitted word in 93\% of trajectories and has post-hit residence one, but decoded BER is 0.01054 rather than the additive value 0.00595; additive acquisition is 95\%.  Binary-state inertia can therefore contribute substantially without reproducing the full saturated-response benefit.

\section{Response alignment and trajectory diagnostics}

\begin{figure}[ht]
\centering
\includegraphics[width=0.82\textwidth]{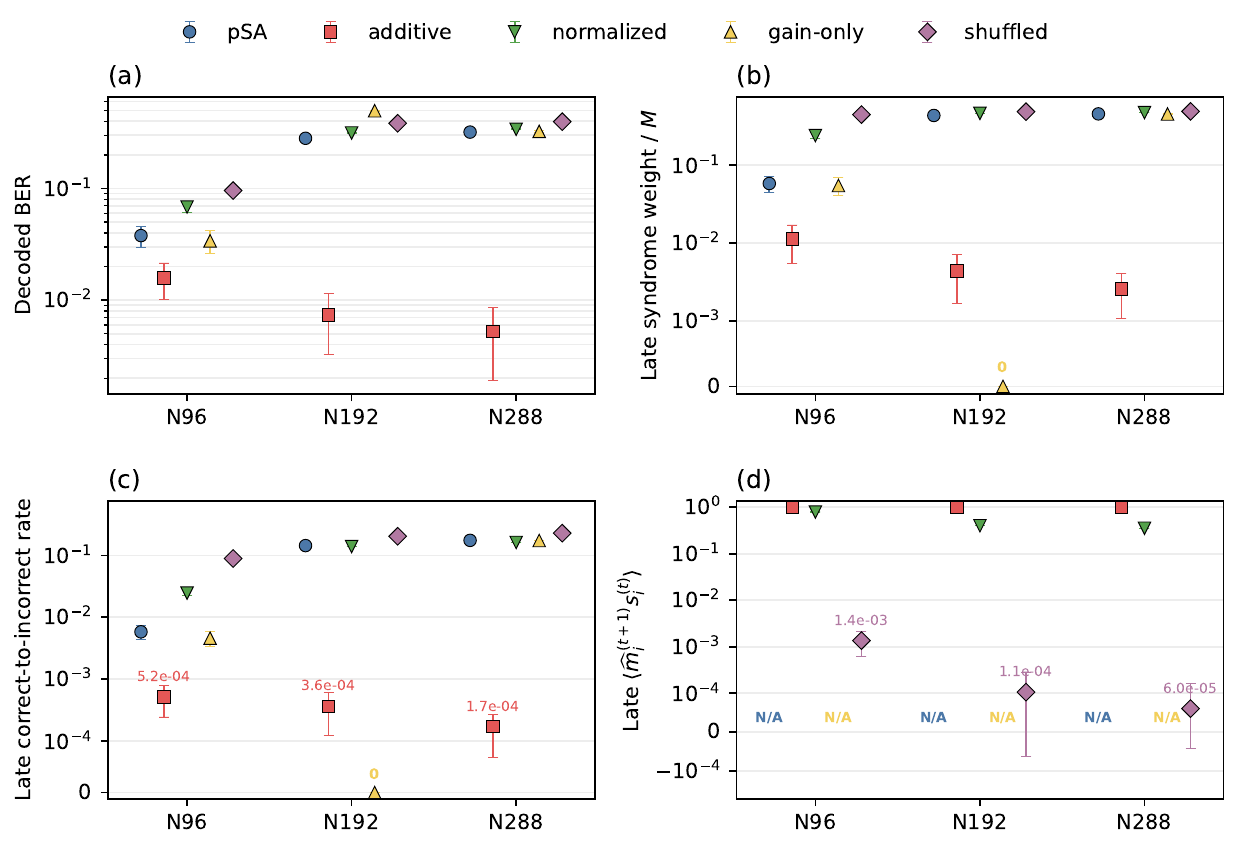}
\caption{Response-alignment and trajectory diagnostics at 2.5 dB.  Points and error bars retain the means and 95\% intervals in the archived summaries.  Panel (a) uses a logarithmic scale; panels (b)--(d) use symmetric-log coordinates with a linear neighborhood of zero, so exact zeros are displayed without a positive plotting floor.  Labeled zero markers denote observed values of exactly zero.  N/A denotes methods for which no delayed response source is supplied and the used-product metric is therefore not applicable.  Panel (c) shows the late correct-to-incorrect rate and panel (d) the retained-response/selected-source product, including held cycles, as defined in this section.  Small nonzero values in panels (c) and (d) are annotated in scientific notation.}
\label{figS:alignment-summary}
\end{figure}

\begin{figure}[ht]
\centering
\includegraphics[width=0.98\textwidth]{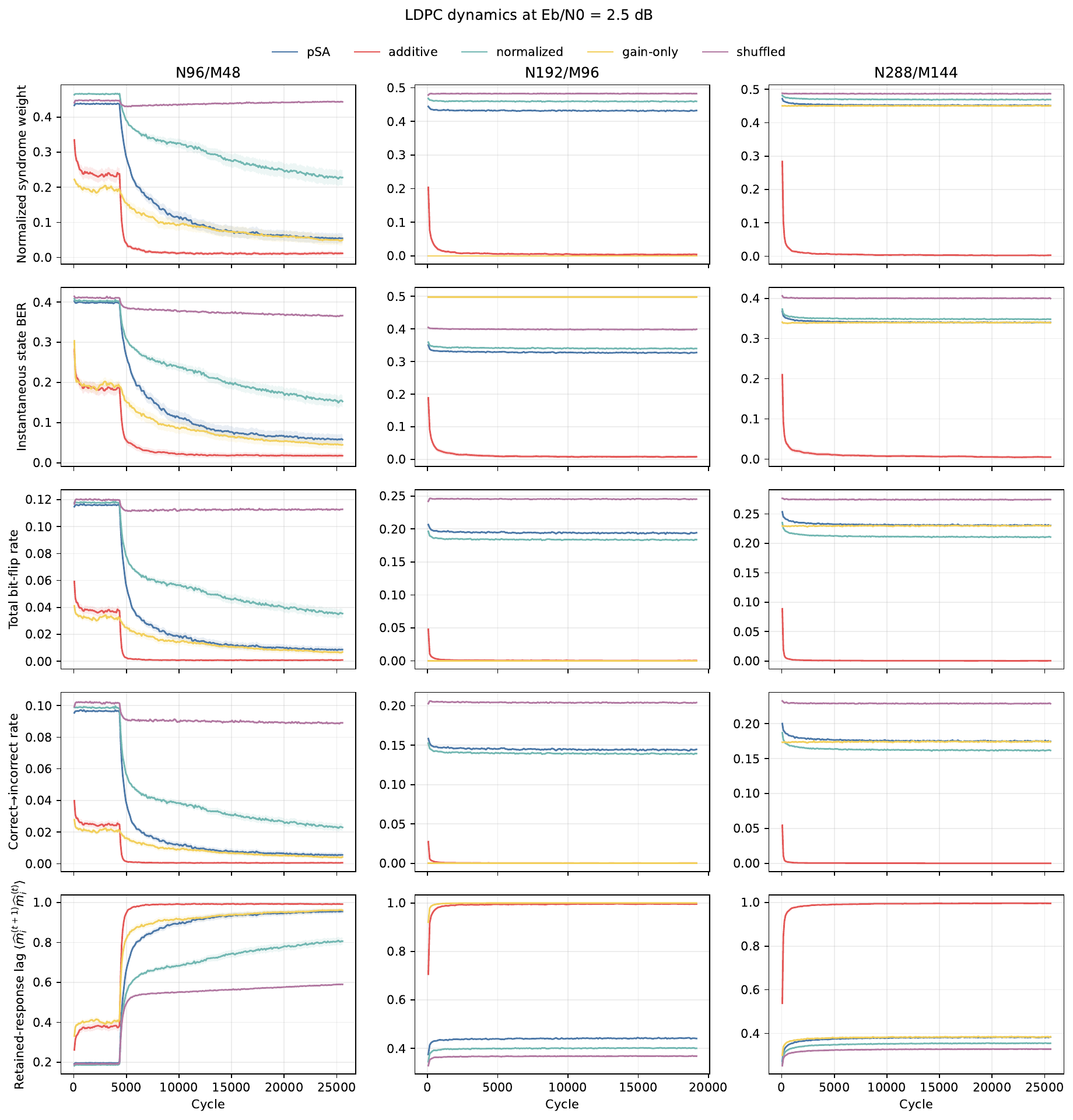}
\caption{Size-resolved trajectories with archived trajectory-level uncertainty bands.  The five rows show normalized syndrome weight, instantaneous state BER, total bit-flip rate, correct-to-incorrect back-flip rate, and the retained-response same-bit lag product $\langle\widehat m_i^{(t+1)}\widehat m_i^{(t)}\rangle$, respectively.  The last row includes held cycles and differs from the selected-source product in Fig.~\ref{figS:alignment-summary}(d) for shuffling.}
\label{figS:alignment-dynamics}
\end{figure}

The trajectory analysis contains 9000 trajectories: three sizes, five rules, three SNRs, and 200 trajectories per cell.  Correctness is always relative to the transmitted word.  Total bit-flip rate counts all reversals per bit-cycle; back-flip rate counts correct-to-incorrect transitions divided by previously correct bit states, including held bits.  Within each trajectory the logger pools numerator and denominator over each 100-cycle bin before division.  Late values pool counts over the final quarter of cycles before taking the ratio; the selected cycle counts align with the bin boundary.  The correct-start diagnostics use 100-cycle bins and cross-code diagnostics use 200-cycle bins, with the same pooled-count definition.  A zero denominator gives NaN, excluded from finite-value summaries.  Figures~\ref{figS:alignment-summary} and \ref{figS:alignment-dynamics} display means over the 200 trajectory-level values, with the archived normal-approximation intervals (mean $\pm1.96$ standard errors); these are not means of cycle-wise ratios or seed-batch intervals.  Separately reported paired contrasts first average within each seed batch and use the ten paired batch means.

For response diagnostics, let $\widehat m_i^{(t)}$ denote the logged retained response at the start of cycle $t$.  On activation, $\widehat m_i^{(t+1)}=q_i^{(t)}$; on a hold it remains $\widehat m_i^{(t)}$.  It equals the stored response $m_i^{(t)}$ for additive, normalized, and shuffled rules; for \psa{} and gain-only it is only an unused diagnostic log, not algorithmic memory.  Same-bit persistence is the uncentered product $\widehat m_i^{(t+1)}\widehat m_i^{(t)}$.  The distinct selected-source product is $\widehat m_i^{(t+1)}s_i^{(t)}$, where $s_i^{(t)}=\widehat m_i^{(t)}$ for additive/normalized rules and $s_i^{(t)}=\widehat m_{\pi_t(i)}^{(t)}$ for shuffling.  The cycle-specific source mapping is fixed before commitment.  Both products include held bit-cycles: in a held cycle the source is recorded diagnostically but is not injected into an update.  Products are summed over bits and cycles and divided by the number of bit-cycles, then averaged over trajectories; late values use the same final-quarter window.  These products are neither Pearson correlations nor cosine similarities.  The selected-source metric is N/A for \psa{} and gain-only, even though their diagnostic same-bit lag product can be computed.

Figure~\ref{figS:alignment-summary} summarizes the key 2.5-dB diagnostics, and Fig.~\ref{figS:alignment-dynamics} provides all size-resolved temporal panels.

\begin{table}[ht]
\caption{Key response-alignment and trajectory values at 2.5 dB.  The late back-flip columns use the pooled-count correct-to-incorrect rate defined in this section.  ``Used product'' denotes the retained-response/selected-source product $\langle\widehat m_i^{(t+1)}s_i^{(t)}\rangle$, including held cycles where the source is diagnostic only, not injected.}
\label{tabS:alignment}
\begin{ruledtabular}
\begin{tabular}{lcccccccc}
$N$ & \multicolumn{3}{c}{Decoded BER} & Add. synd./$M$ & \multicolumn{3}{c}{Late back-flip rate} & Used product \\
 & Add. & \psa{} & Shuf. & & Add. & \psa{} & Shuf. & Shuf. \\
\hline
96  & 0.01583 & 0.03786 & 0.09583 & 0.01130 & 0.000515 & 0.005815 & 0.088996 & 0.001360 \\
192 & 0.007370 & 0.28172 & 0.38367 & 0.004421 & 0.000363 & 0.14402 & 0.20390 & 0.000106 \\
288 & 0.005226 & 0.32009 & 0.39649 & 0.002601 & 0.000170 & 0.17492 & 0.22871 & 0.000060 \\
\end{tabular}
\end{ruledtabular}
\end{table}

The derangement has no self-links.  Across the evaluated trajectories, the maximum absolute discrepancies between original and shuffled delayed-response distributions are $1.16\times10^{-17}$ for the mean and $5.90\times10^{-18}$ for the second moment.  The paired shuffled-minus-additive BER differences at 2.5 dB are 0.0800 $[0.0747,0.0853]$, 0.3763 $[0.3704,0.3822]$, and 0.3913 $[0.3865,0.3960]$ for $N=96,192,288$, respectively.  Brackets are 95\% intervals across ten paired seed-batch means.

\section{Acquisition and post-acquisition stability}

Default initialization and correct-start retention are distinct interventions.  The default study begins from the implementation's all-zero bit state and zero response state; because the transmitted codeword is randomly generated, the initial valid codeword is generally not the transmitted word.  The correct-start study places all methods at the transmitted word but initializes response memory to zero, thereby asking how the target is retained before a matching history has accumulated.

\begin{figure}[ht]
\centering
\includegraphics[width=0.94\textwidth]{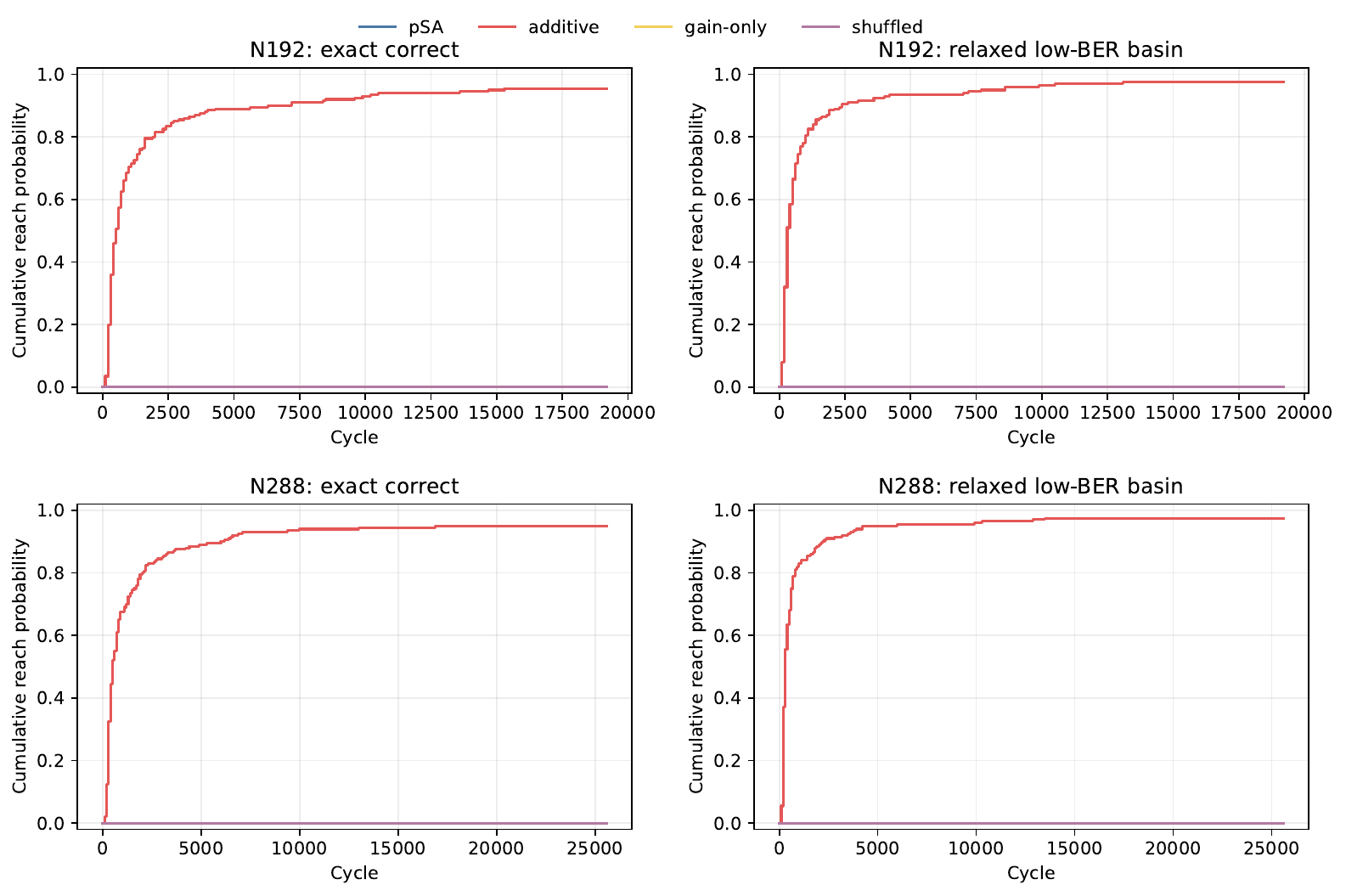}
\caption{Default all-zero-initialization first-passage analysis.  Nonreaching trajectories remain censored; reach fractions are reported separately rather than assigned an artificial passage time.}
\label{figS:firstpassage}
\end{figure}

\begin{figure}[ht]
\centering
\includegraphics[width=0.94\textwidth]{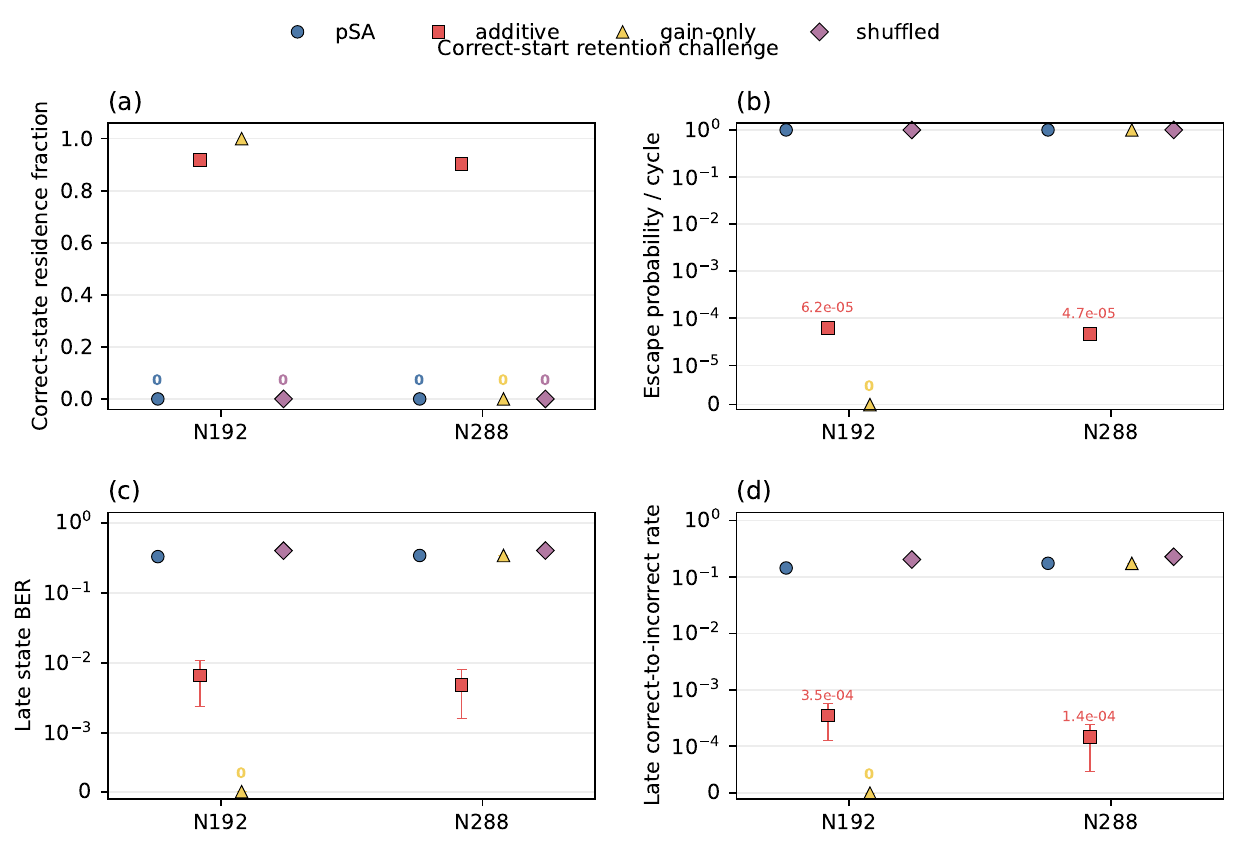}
\caption{Correct-start retention challenge.  Every method begins at the transmitted codeword with zero response memory.  Points retain the archived values and existing intervals.  Labeled zero markers denote observed values of exactly zero; panels (b)--(d) use symmetric-log coordinates with a linear neighborhood of zero and no substituted positive floor.  Positive values below $10^{-3}$ are annotated in scientific notation.  Panel (d) is the late correct-to-incorrect back-flip rate, normalized by previously correct bit states including held bits.  All four metrics are defined for every method in this intervention.  This intervention must not be read as a default-initialization decoding success rate.}
\label{figS:retention}
\end{figure}

\begin{table}[ht]
\caption{Default-initialization acquisition and post-hit residence at 2.5 dB; 200 trajectories per method and size.}
\label{tabS:acquisition-stability}
\begin{ruledtabular}
\begin{tabular}{llccccc}
$N$ & Method & Decoded BER & Correct reach & Median / mean FPT & Residence & Escape/cycle \\
\hline
192 & \psa{} & 0.2892 & 0 & censored & -- & -- \\
192 & additive & 0.00680 & 0.955 & 444 / 1335 & 0.999986 & $5.28\times10^{-6}$ \\
192 & gain-only & 0.4972 & 0 & censored & -- & -- \\
192 & shuffled & 0.3869 & 0 & censored & -- & -- \\
288 & \psa{} & 0.3245 & 0 & censored & -- & -- \\
288 & additive & 0.00595 & 0.950 & 425 / 1261 & 0.999982 & $3.46\times10^{-6}$ \\
288 & gain-only & 0.3225 & 0 & censored & -- & -- \\
288 & shuffled & 0.3961 & 0 & censored & -- & -- \\
\end{tabular}
\end{ruledtabular}
\end{table}

In the correct-start challenge, additive residence is 0.9176 at $N=192$ and 0.9020 at $N=288$; conditional return probabilities after escape are approximately 0.9587 and 0.9628.  \psa{} and shuffled memory leave immediately and have zero residence at the sampled resolution.  Gain-only has residence one at $N=192$ but zero at $N=288$, further demonstrating that residence without channel-consistent acquisition is not a transferable success mechanism.

The $N=192$ gain-only default-initialization trajectories freeze at zero syndrome and zero late flip rate, but late BER is 0.497 and normalized channel alignment is 0.0074.  The additive values are 0.0066 and 0.695.  This is the valid-wrong-codeword counterexample used in the main text.

\subsection{Initialization robustness}

We tested whether the acquisition result above depends on starting from the valid all-zero codeword.  The same representative $N=192$ and 288 matrices, 2.5-dB channel, fixed matched-control parameter sets, four response rules, ten paired seed batches, and 200 trajectories per method and initialization were retained.  No parameter was retuned.  A random start draws each stored bit independently and equiprobably from zero or one using a dedicated initialization stream; a channel-hard start sets bit one exactly when the received BPSK sample is negative.  Within a paired trajectory all methods receive the same channel realization and initial state, and the response state is zero.

\begin{table}[ht]
\caption{Initial-state diagnostics for the alternative-initialization study.  Syndrome is normalized by the number of checks.}
\label{tabS:init-diagnostics}
\begin{ruledtabular}
\begin{tabular}{llccc}
$N$ & Initialization & Initial BER & Syndrome & Channel alignment \\
\hline
192 & random & 0.5017 & 0.5007 & 0.00015 \\
192 & channel hard decision & 0.09237 & 0.3564 & 0.7691 \\
288 & random & 0.4996 & 0.4971 & 0.00147 \\
288 & channel hard decision & 0.09297 & 0.3522 & 0.8100 \\
\end{tabular}
\end{ruledtabular}
\end{table}

\begin{figure}[ht]
\centering
\includegraphics[width=0.98\textwidth]{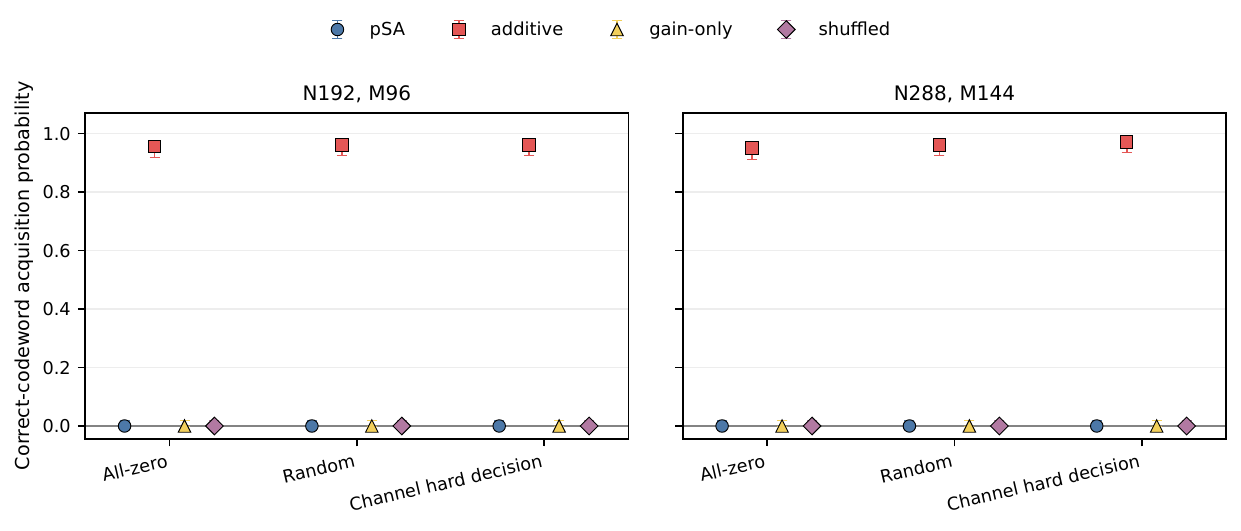}
\caption{Correct-codeword acquisition under all-zero, random, and channel-hard-decision initialization.  Each cell contains 200 trajectories at 2.5 dB under the fixed matched-control parameter set.  Markers on the zero baseline denote observed acquisition proportions of exactly zero, rather than missing or inapplicable data.  Error bars are the existing binomial Wilson intervals for the displayed acquisition proportion; the paired method contrasts are reported separately in the text.}
\label{figS:init-acquisition}
\end{figure}

\begin{table}[ht]
\caption{Alternative-initialization decoding and acquisition outcomes.  First-passage time (FPT) is summarized only for additive trajectories that reach the transmitted codeword.}
\label{tabS:init-outcomes}
\scriptsize
\begin{ruledtabular}
\begin{tabular}{llcccccccc}
& & \multicolumn{4}{c}{Correct reach} & \multicolumn{2}{c}{Decoded BER} & \multicolumn{2}{c}{Additive post-hit} \\
$N$ & Initialization & \psa{} & Add. & Gain & Shuf. & \psa{} & Add. & FPT median/mean & Residence \\
\hline
192 & random & 0 & 0.96 & 0 & 0 & 0.2831 & 0.00583 & 436.5/1431 & 0.999231 \\
192 & channel hard & 0 & 0.96 & 0 & 0 & 0.2836 & 0.00591 & 500.5/1399 & 0.999986 \\
288 & random & 0 & 0.96 & 0 & 0 & 0.3237 & 0.00458 & 548/1337 & 0.999996 \\
288 & channel hard & 0 & 0.97 & 0 & 0 & 0.3208 & 0.00345 & 457/1678 & 0.999962 \\
\end{tabular}
\end{ruledtabular}
\end{table}

The additive-minus-matched-\psa{} acquisition differences and 95\% intervals are 96.0 $[92.7,99.3]$ and 96.0 $[93.2,98.8]$ percentage points at $N=192$, and 96.0 $[91.6,100.4]$ and 97.0 $[94.0,100.0]$ percentage points at $N=288$, for random and channel-hard starts, respectively.  These are unbounded paired-$t$ intervals over the same ten prespecified seed-batch differences used by the matched trajectory framework.  They are not clipped to the probability-difference range; the 100.4\% upper endpoint is therefore a finite-sample interval endpoint, not a probability estimate.  We retain this construction for consistency rather than selecting a different interval method for this validation alone.

\begin{figure}[ht]
\centering
\includegraphics[width=0.98\textwidth]{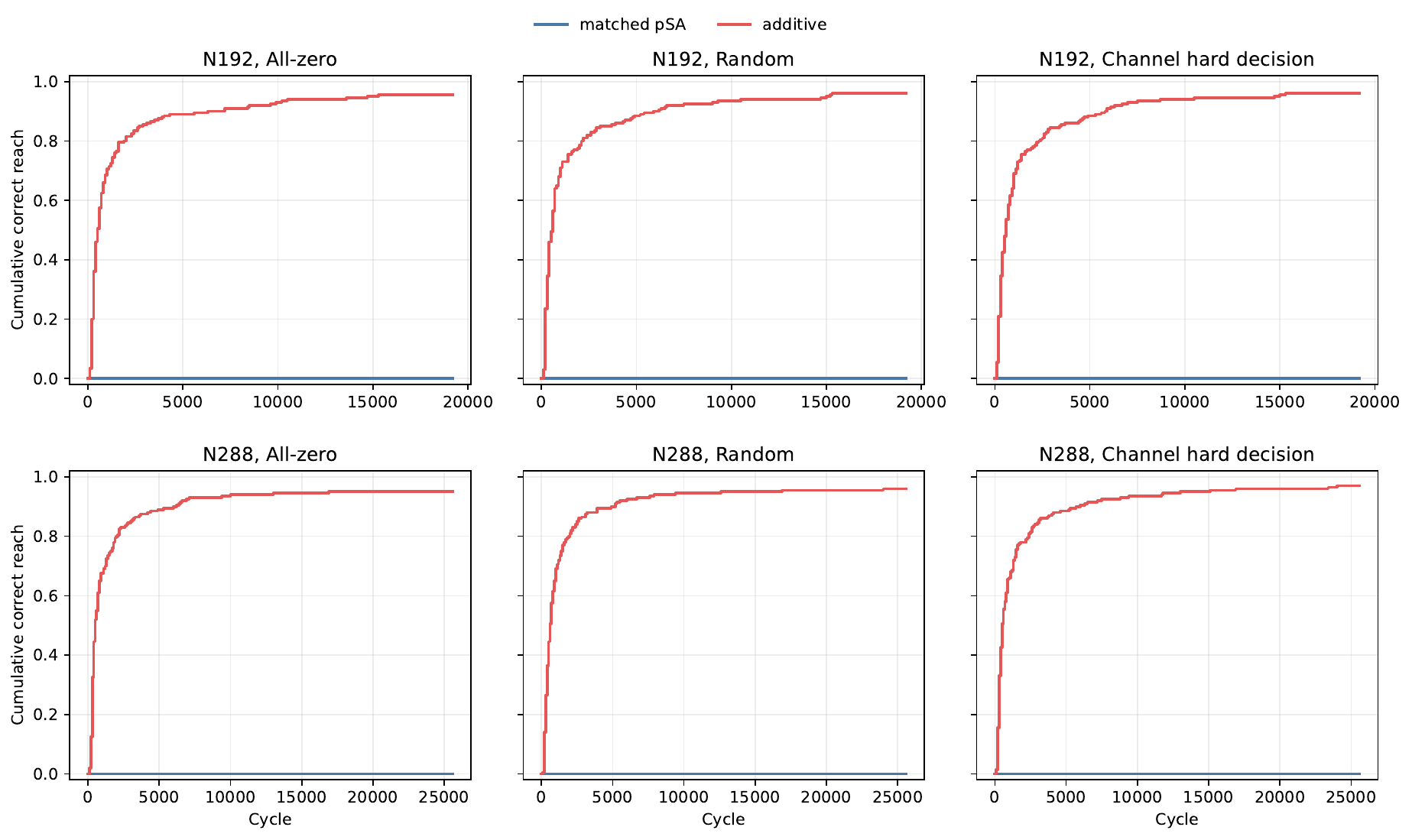}
\caption{Censored cumulative correct-codeword reach under the three initializations.  Nonreaching trajectories are not assigned the terminal cycle.  Because matched \psa{} does not reach the correct word in any cell, an additive-versus-\psa{} conditional FPT difference is undefined rather than imputed.}
\label{figS:init-firstpassage}
\end{figure}

Across the four alternative-initialization cells, additive late instantaneous BER is 0.0038--0.0058, compared with 0.328--0.341 for matched \psa{}.  Additive post-hit residence remains above 0.9992.  Gain-only and shuffled memory acquire the transmitted codeword in zero trajectories in every cell.  Thus the acquisition advantage is not an artifact of the valid all-zero start for the tested representative matrices, block lengths, SNR, and initialization rules; broader initial-state, channel, and code-ensemble dependence is not established.

\section{Cross-code transfer robustness}

The fixed-transfer evaluation has three distinct arms.  The additive and pSA-specific packages were independently selected on the representative matrix for each block length and then transferred intact to the same C00--C09 matrices without per-code retuning.  This is a package-level performance comparison: each independently selected package retains its own readout and other recorded fields, as documented in Table~\ref{tabS:parameters}.  The matched $\lambda=0$ ablation instead uses the additive package with only its response coefficient removed, so its readout and all nonmemory parameters remain identical to the additive arm.  All arms use the same code instances, SNRs, stochastic seed design, channel generation, and initialization.  Main Fig.~5 gives the three-way code-level BERs.

\begin{table}[ht]
\caption{Median relative error-rate reduction of additive fixed transfer against the two fixed-transfer comparators.  Brackets are 95\% percentile intervals from 20,000 bootstrap resamples of the ten independent code realizations.}
\label{tabS:transfer-effects}
\begin{ruledtabular}
\begin{tabular}{clcc}
$N$ & Comparator & BER reduction (\%) & FER reduction (\%) \\
\hline
96  & pSA-specific & 35.46 $[32.29,38.83]$ & 42.80 $[38.84,45.12]$ \\
192 & pSA-specific & 76.61 $[74.81,77.17]$ & 81.81 $[79.71,82.84]$ \\
288 & pSA-specific & 81.10 $[80.66,81.84]$ & 82.56 $[82.08,83.00]$ \\
96  & matched $\lambda=0$ & 59.90 $[58.37,61.14]$ & 62.70 $[62.25,65.15]$ \\
192 & matched $\lambda=0$ & 97.46 $[97.29,97.56]$ & 94.22 $[93.42,94.53]$ \\
288 & matched $\lambda=0$ & 97.61 $[97.56,97.70]$ & 92.85 $[92.63,93.13]$ \\
\end{tabular}
\end{ruledtabular}
\end{table}

\begin{figure}[ht]
\centering
\includegraphics[width=0.88\textwidth]{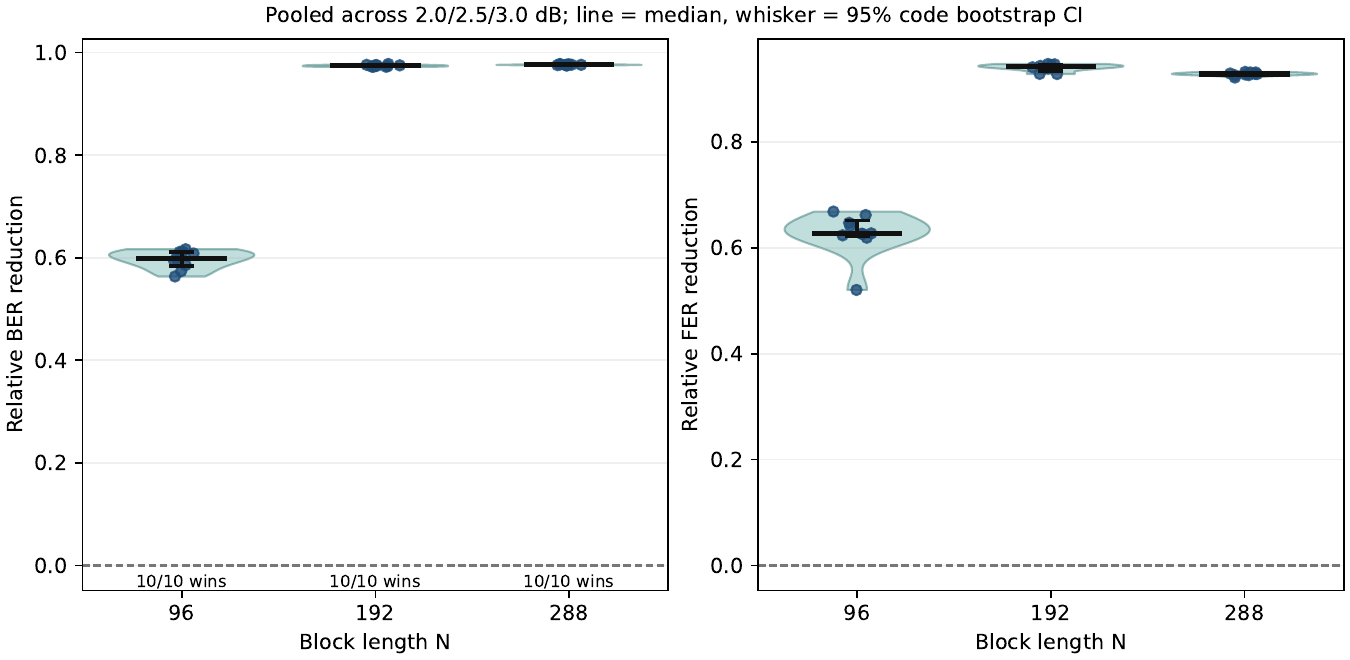}
\caption{Code-level BER and FER changes relative to the matched $\lambda=0$ ablation under fixed size-specific parameter transfer.  These reductions are causal rule-ablation effects; the pSA-specific fixed-transfer comparison is reported in main Fig.~5 and Table~\ref{tabS:transfer-effects}.}
\label{figS:crosscode-summary}
\end{figure}

\begin{figure}[ht]
\centering
\includegraphics[width=0.98\textwidth]{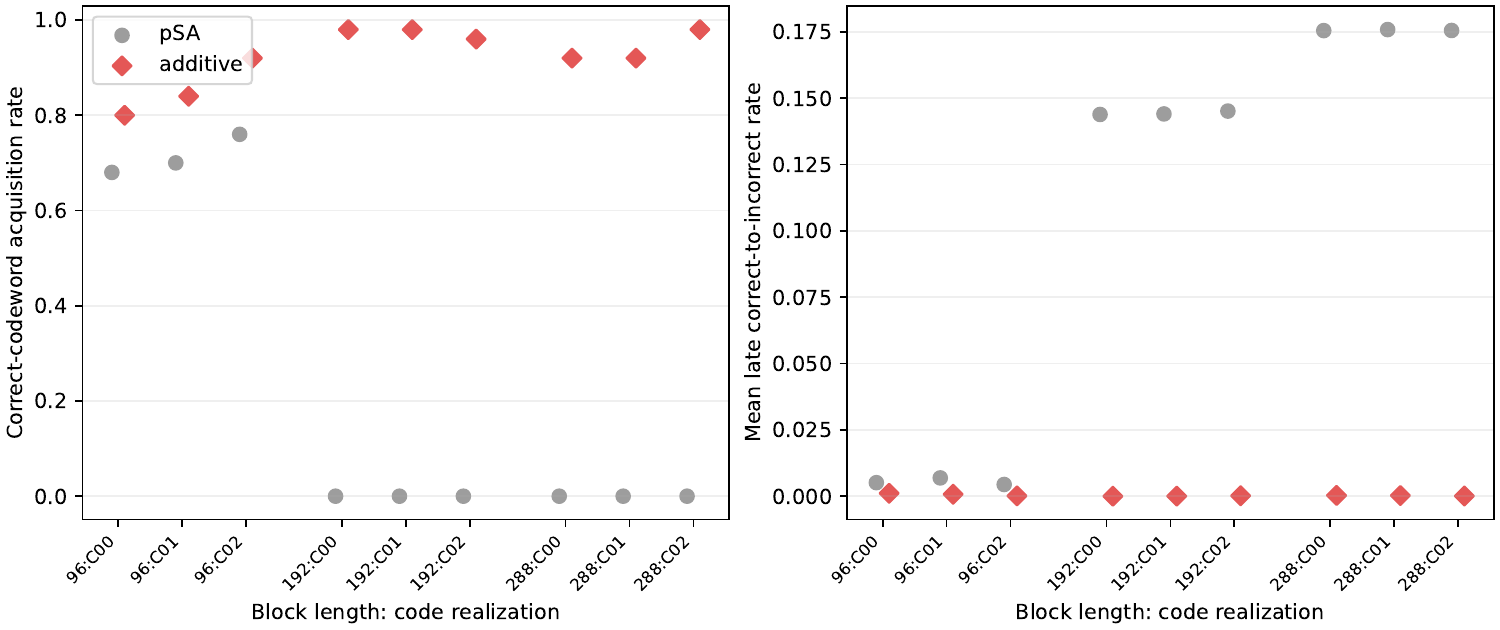}
\caption{Mechanism sampling on nine independent code realizations.  The right panel is the trajectory-mean late correct-to-incorrect back-flip rate, with previously correct bit states (including held bits) as the denominator.  The general transferable descriptor is post-acquisition stability: at $N=96$, faster return can matter even when escape frequency alone is not lower.}
\label{figS:crosscode-mechanism}
\end{figure}

The additive fixed-transfer package has lower BER and FER than the pSA-specific fixed-transfer package for all 30 code realizations.  The pSA-specific comparator is the package-level performance baseline; the larger matched-ablation effects isolate the response-rule change under additive-selected nonmemory parameters and readout and are not independently optimized margins.  Figure~\ref{figS:crosscode-summary} retains the matched-ablation size summary for that causal comparison, while Fig.~\ref{figS:crosscode-mechanism} summarizes trajectory diagnostics for nine sampled matrices.  Additive dynamics have higher correct-basin acquisition and lower late correct-to-incorrect back-flip rate in all nine.

\clearpage

\begingroup

\normalsize
\setlength{\tabcolsep}{4pt}          
\renewcommand{\arraystretch}{1.10}
\makeatletter
\let\@arstrut\@arstrut@org
\let\@arstrut@hook\@empty
\def\table@hook{\normalsize
  \setbox\strutbox\hbox{\vrule height9.8pt depth4.2pt width0pt}}
\makeatother
\setlength{\LTpre}{0pt}
\setlength{\LTpost}{0pt}

\begin{longtable}{lrrrrr}
	
	\caption{Code-level additive and matched-$\lambda=0$ outcomes for the 30 transferred code realizations.
		BER and FER are pooled over the three SNRs; Red. denotes the additive BER reduction relative to
		the matched arm. The pSA-specific results are shown in main Fig.~5.}
	\label{tabS:codes}
	\\
	
	\toprule
	
	
	\multicolumn{6}{c}{\textbf{(a)} $N=96$}
	\\[2pt]

	\cmidrule(lr){1-6}
	
	Code
	& Matched BER
	& Add. BER
	& Red. (\%)
	& Matched FER
	& Add. FER
	\\[2pt]
	
	\midrule
	
	C00 & 0.03576 & 0.01561 & 56.35 & 0.3740 & 0.1793 \\[3.5pt]
	C01 & 0.03420 & 0.01339 & 60.84 & 0.3047 & 0.1030 \\[3.5pt]
	C02 & 0.03487 & 0.01414 & 59.46 & 0.3250 & 0.1213 \\[3.5pt]
	C03 & 0.03574 & 0.01452 & 59.36 & 0.3190 & 0.1193 \\[3.5pt]
	C04 & 0.03337 & 0.01383 & 58.57 & 0.3047 & 0.1147 \\[3.5pt]
	C05 & 0.03570 & 0.01416 & 60.34 & 0.3133 & 0.1193 \\[3.5pt]
	C06 & 0.03376 & 0.01312 & 61.14 & 0.3017 & 0.1000 \\[3.5pt]
	C07 & 0.03399 & 0.01449 & 57.38 & 0.3050 & 0.1137 \\[3.5pt]
	C08 & 0.03608 & 0.01401 & 61.18 & 0.3213 & 0.1153 \\[3.5pt]
	C09 & 0.03385 & 0.01298 & 61.66 & 0.3023 & 0.1067 \\[3.5pt]
	
	\addlinespace[6pt]
	\midrule
	\addlinespace[3pt]
	
	
	\multicolumn{6}{c}{\textbf{(b)} $N=192$}
	\\[2pt]

	\cmidrule(lr){1-6}
	
	Code
	& Matched BER
	& Add. BER
	& Red. (\%)
	& Matched FER
	& Add. FER
	\\[2pt]
	
	\midrule
	
	C00 & 0.28130 & 0.00763 & 97.29 & 0.9997 & 0.0603 \\[3.5pt]
	C01 & 0.28173 & 0.00774 & 97.25 & 1.0000 & 0.0713 \\[3.5pt]
	C02 & 0.28082 & 0.00686 & 97.56 & 1.0000 & 0.0563 \\[3.5pt]
	C03 & 0.28261 & 0.00675 & 97.61 & 0.9997 & 0.0547 \\[3.5pt]
	C04 & 0.28251 & 0.00783 & 97.23 & 1.0000 & 0.0717 \\[3.5pt]
	C05 & 0.28336 & 0.00628 & 97.78 & 1.0000 & 0.0527 \\[3.5pt]
	C06 & 0.28149 & 0.00699 & 97.52 & 1.0000 & 0.0610 \\[3.5pt]
	C07 & 0.28260 & 0.00754 & 97.33 & 1.0000 & 0.0590 \\[3.5pt]
	C08 & 0.28201 & 0.00701 & 97.51 & 1.0000 & 0.0533 \\[3.5pt]
	C09 & 0.28216 & 0.00733 & 97.40 & 1.0000 & 0.0567 \\[3.5pt]
	
	\pagebreak[4]
	
	\addlinespace[6pt]
	\midrule
	\addlinespace[3pt]
	
	
	\multicolumn{6}{c}{\textbf{(c)} $N=288$}
	\\[2pt]

	\cmidrule(lr){1-6}
	
	Code
	& Matched BER
	& Add. BER
	& Red. (\%)
	& Matched FER
	& Add. FER
	\\[2pt]
	
	\midrule
	
	C00 & 0.32279 & 0.00773 & 97.61 & 1.0000 & 0.0707 \\[3.5pt]
	C01 & 0.32230 & 0.00798 & 97.52 & 1.0000 & 0.0783 \\[3.5pt]
	C02 & 0.32287 & 0.00768 & 97.62 & 1.0000 & 0.0737 \\[3.5pt]
	C03 & 0.32326 & 0.00733 & 97.73 & 1.0000 & 0.0687 \\[3.5pt]
	C04 & 0.32218 & 0.00761 & 97.64 & 1.0000 & 0.0717 \\[3.5pt]
	C05 & 0.32234 & 0.00785 & 97.57 & 1.0000 & 0.0713 \\[3.5pt]
	C06 & 0.32247 & 0.00811 & 97.49 & 1.0000 & 0.0740 \\[3.5pt]
	C07 & 0.32295 & 0.00743 & 97.70 & 1.0000 & 0.0687 \\[3.5pt]
	C08 & 0.32223 & 0.00715 & 97.78 & 1.0000 & 0.0677 \\[3.5pt]
	C09 & 0.32262 & 0.00777 & 97.59 & 1.0000 & 0.0727 \\[3.5pt]
	
	\bottomrule
	
\end{longtable}

\endgroup

\clearpage

All 30 rows favor the additive response rule in BER and FER against the matched ablation.  Code labels C00--C09 are presentation labels linked to matrix hashes in the archived metadata.  With 20,000 code-level bootstrap resamples, the median relative BER reduction at $N=96$ is 59.90\%, with interval $[58.37,61.14]\%$.  The corresponding values are 97.46\% $[97.29,97.56]\%$ at $N=192$ and 97.61\% $[97.56,97.70]\%$ at $N=288$.  These matched-ablation values are not substituted for the pSA-specific fixed-transfer effects in Table~\ref{tabS:transfer-effects}.

\section{Cross-problem boundary controls}

These experiments test for a generic endpoint advantage from temporal state, but they do not constitute a matched causal test of the additive LDPC rule across problems.  MAX-CUT outcomes are co-optimized, whereas random 2-SAT uses a matched-$k_w$ finite-response $\rho$ control.  They are not additional applications used to support the LDPC mechanism, and their objective scales are not compared directly with BER or FER.

\subsection{MAX-CUT}

For an undirected graph $G=(V,E)$ with adjacency matrix $A$, the cut represented by $x_i\in\{-1,+1\}$ is
\begin{equation}
C(\bm x)=\frac{1}{2}\sum_{(i,j)\in E}(1-x_i x_j).
\label{eqS:maxcutvalue}
\end{equation}
The implementation minimizes
\begin{align}
\mathcal E_{\mathrm{MC}}(\bm x)
&=k_w\sum_{(i,j)\in E}x_i x_j
=k_w[|E|-2C(\bm x)],\notag\\
F_{i,\mathrm{MC}}&=-k_w\sum_j A_{ij}x_j.
\label{eqS:maxcutfield}
\end{align}
Endpoint quality is the mean cut normalized by $|E|$.  These experiments use their own tuning objectives and cannot be combined numerically with BER.

\begin{figure}[ht]
\centering
\includegraphics[width=0.88\textwidth]{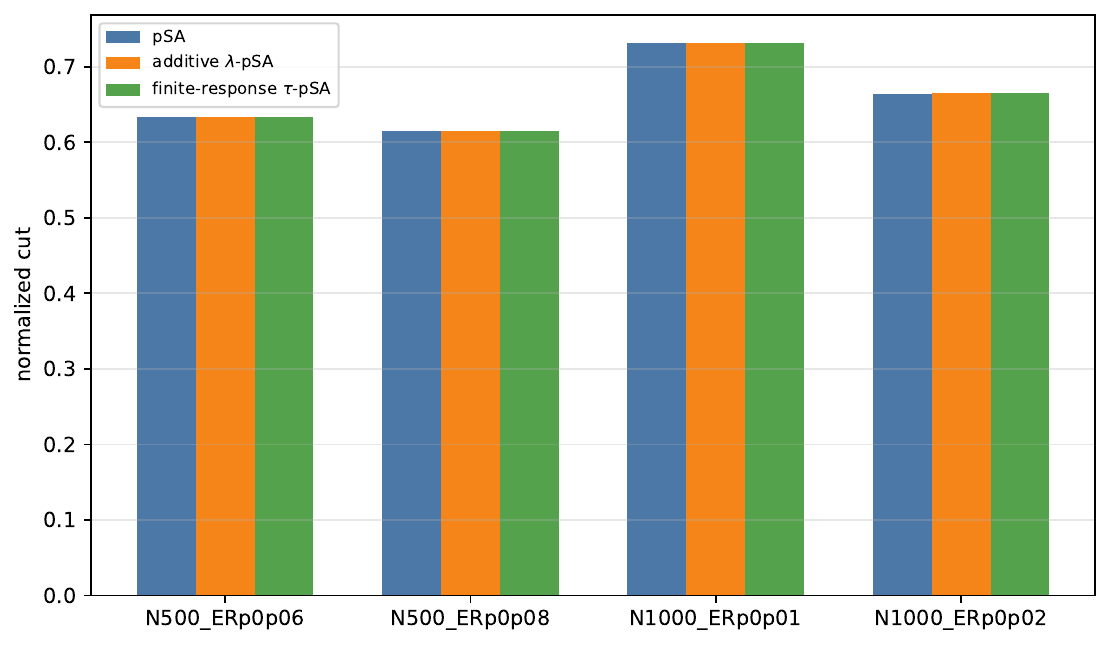}
\caption{Evaluated MAX-CUT endpoint quality.  Differences are small on the displayed scale and are reported as co-optimized outcomes.}
\label{figS:maxcutquality}
\end{figure}

\begin{figure}[ht]
\centering
\includegraphics[width=0.88\textwidth]{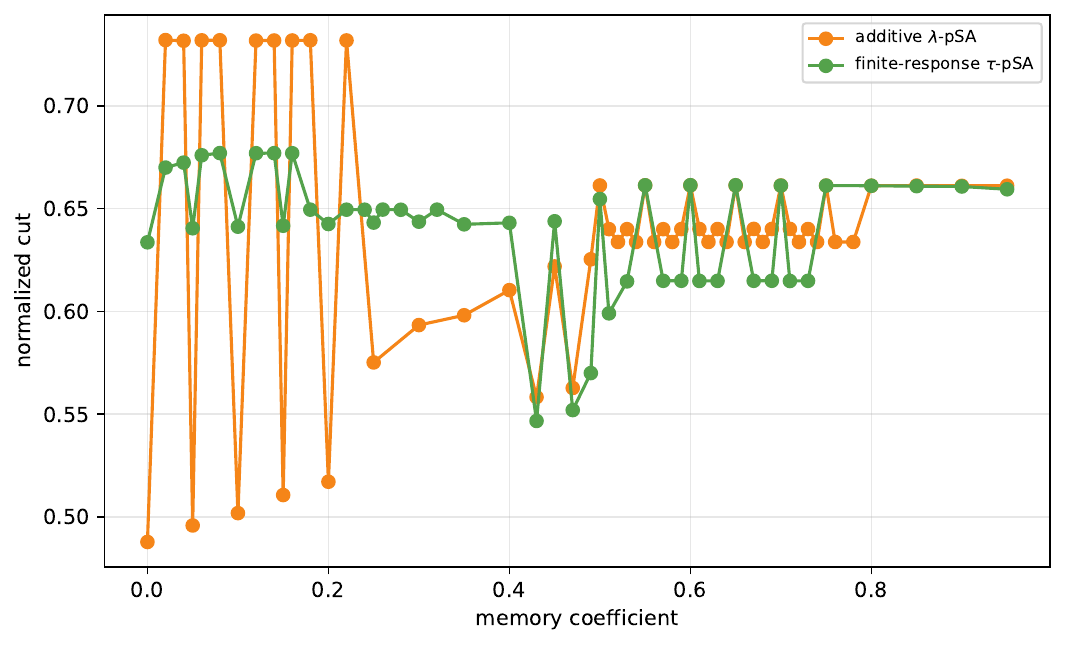}
\caption{MAX-CUT memory sweeps.  Selected endpoints can occur at the memoryless boundary, so a mode label alone is not evidence for active response memory.}
\label{figS:maxcutsweep}
\end{figure}

\begin{table}[ht]
\caption{Selected MAX-CUT outcomes.  Gain is the mean normalized-cut change relative to \psa{}.}
\label{tabS:maxcut}
\begin{ruledtabular}
\begin{tabular}{lccc}
Condition & Best labeled mode & Coefficient & Gain (\%) \\
\hline
$N=1000$, ER $p=0.01$ & finite response & $\rho=0$ & 0.184 \\
$N=1000$, ER $p=0.02$ & finite response & $\rho=0.12$ & 0.137 \\
$N=500$, ER $p=0.06$ & finite response & $\rho=0.22$ & 0.030 \\
$N=500$, ER $p=0.08$ & additive & $\lambda=0.57$ & 0.079 \\
\end{tabular}
\end{ruledtabular}
\end{table}

The broader $N=500$ structural campaign covers 14 density/topology conditions, ten graph seeds, and 100 trials per graph.  Only 5 of 14 mean-quality winners use a nonzero response coefficient, and all apparent mean normalized-cut gains remain below 0.2\%.  The evidence is therefore marginal or conditional.

\subsection{Random 2-SAT matched-\texorpdfstring{$k_w$}{kw} control}

A clause $c=(i,a;j,b)$ contains two signed literals, with $a,b\in\{-1,+1\}$ and literal $(i,a)$ true when $a x_i=+1$.  Its exact violation indicator is
\begin{equation}
V_c(\bm x)=\frac{(1-a x_i)(1-b x_j)}{4}.
\label{eqS:clauseviolation}
\end{equation}
The implemented objective and local field are
\begin{align}
\mathcal E_{\mathrm{2SAT}}(\bm x)
&=k_w\sum_c V_c(\bm x)
=k_w\!\left(C_0+\sum_i h_i x_i+\sum_{i<j}J_{ij}x_i x_j\right),\notag\\
F_{i,\mathrm{2SAT}}
&=-k_w\left(h_i+\sum_jJ_{ij}x_j\right).
\label{eqS:2satfield}
\end{align}
The implementation stores symmetric $J$ with zero diagonal.  Each clause adds $1/4$ to $C_0$, $-a/4$ and $-b/4$ to its linear coefficients, and $ab/4$ to both $J_{ij}$ and $J_{ji}$.  The upper-triangular energy sum counts each pair once; the symmetric row gives the local field.  Consequently, $\mathcal E_{\mathrm{2SAT}}/k_w$ is exactly the unsatisfied-clause count.

\begin{figure}[ht]
\centering
\includegraphics[width=0.84\textwidth]{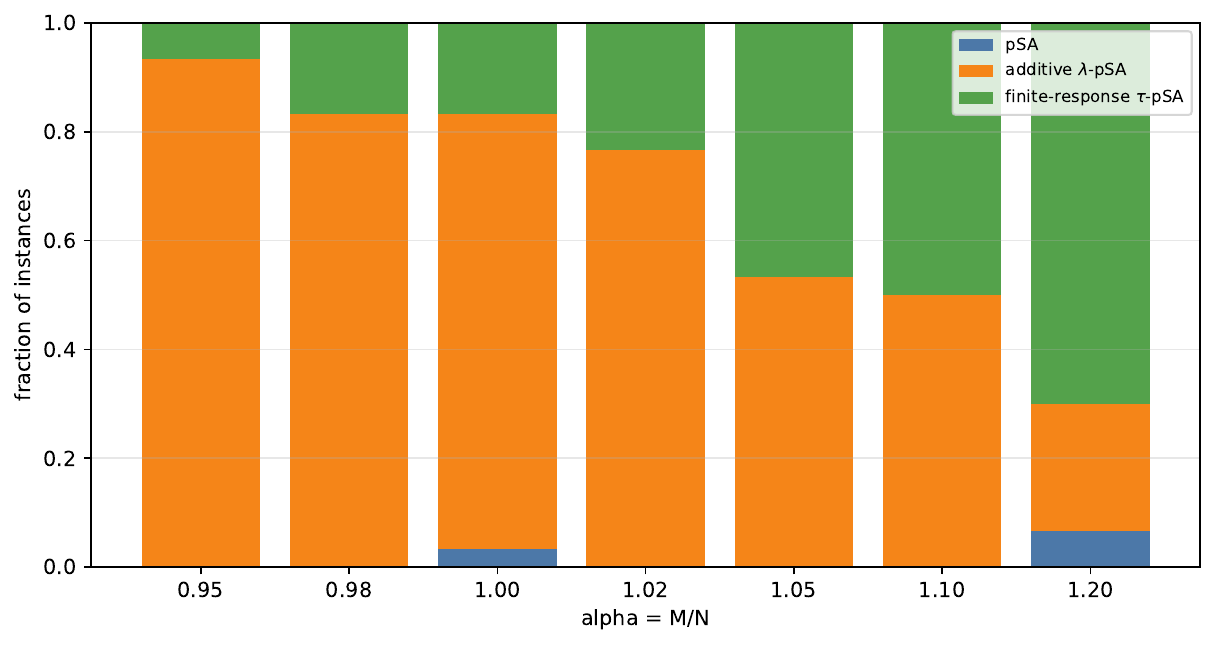}
\caption{Mode distribution from the initial jointly optimized 2-SAT study.  This result is descriptive because $k_w$ and response memory were coupled in optimization.}
\label{figS:initial-dist}
\end{figure}

\begin{figure}[ht]
\centering
\includegraphics[width=0.84\textwidth]{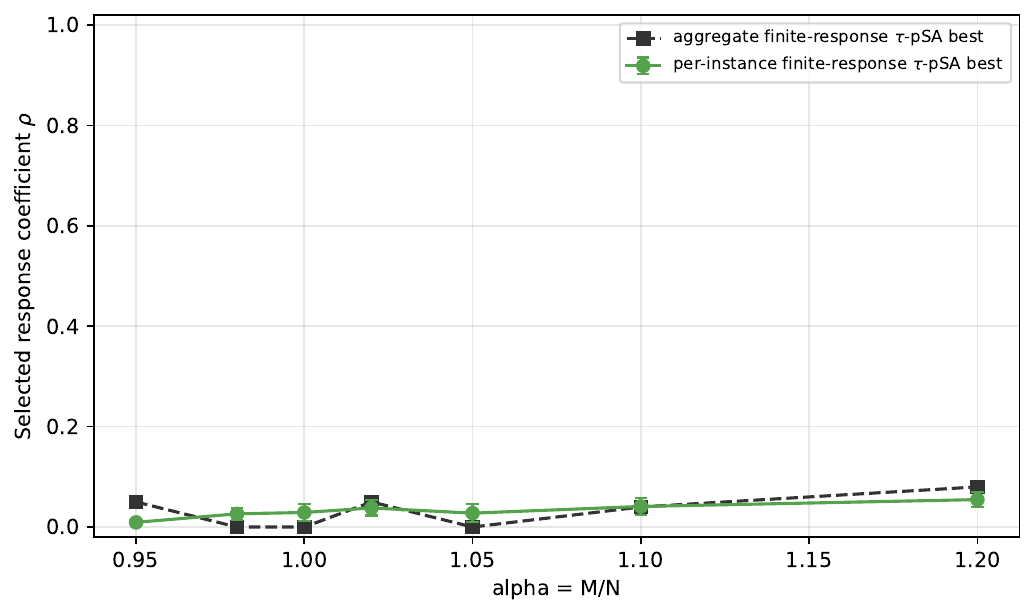}
\caption{Selected response coefficient $\rho$ versus clause density in the initial jointly optimized study.  The apparent density trend is superseded by the matched-$k_w$ causal control.}
\label{figS:initial-alpha}
\end{figure}

The initial exploratory study allowed the optimizer to co-select clause weight $k_w$ and response coefficient $\rho$.  Figures~\ref{figS:initial-dist} and \ref{figS:initial-alpha} retain those observations for transparency, but they are not evidence for an independent response-memory effect.

The matched-$k_w$ control fixes $N=500$, $\alpha=1.20$, the same 30 formulas (13 satisfiable, 17 unsatisfiable), 100 paired trials per formula and cell, and 3000 cycles.  It evaluates $k_w\in\{4,8,12,16,20\}$ and $\rho\in\{0,0.02,0.04,0.06,0.08,0.12\}$ with all remaining parameters fixed.  Table~\ref{tabS:2satgrid} gives the all-formula mean selected unsatisfied-clause rate.  In every row, $\rho=0$ is best; the global minimum is $k_w=8,\rho=0$.  Rows $k_w=12,16,20$ are numerically identical at the displayed precision, excluding a hidden improvement at the earlier $k_w$ search boundary.

\begin{figure}[ht]
\centering
\includegraphics[width=0.90\textwidth]{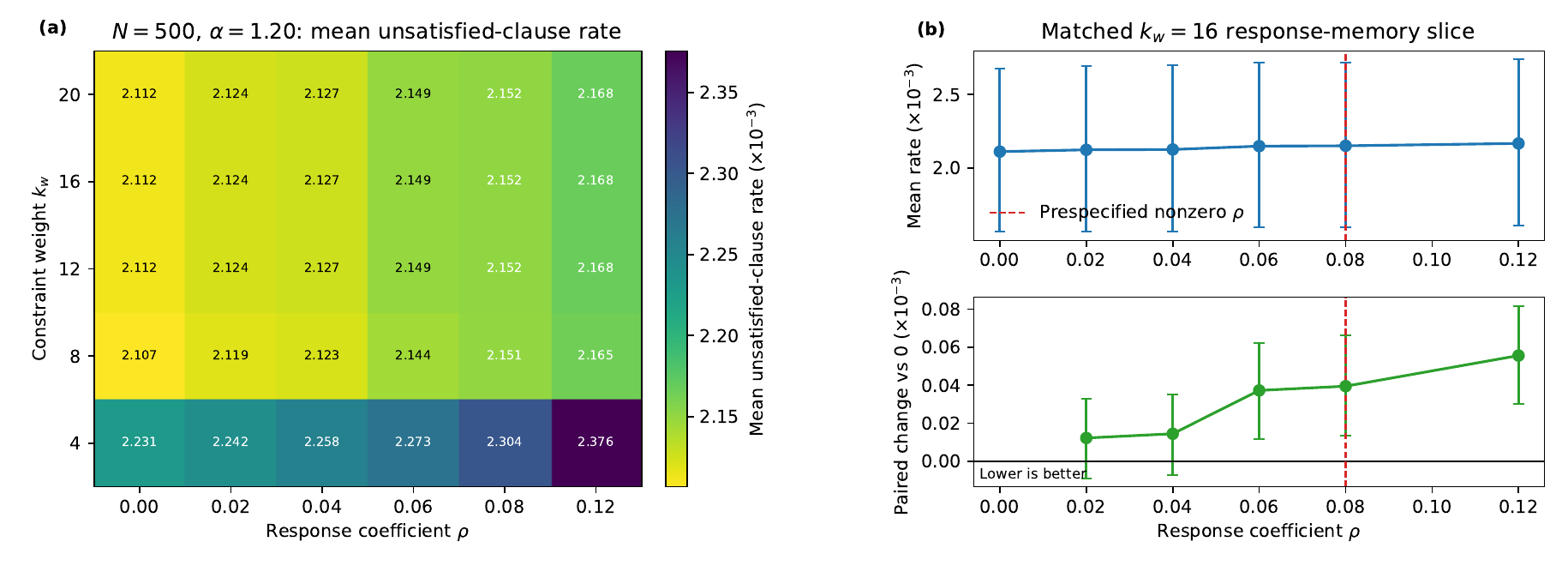}
\caption{Matched-$k_w$ random 2-SAT response-memory control.  Nonzero $\rho$ causes an approximately monotone degradation for every $k_w$.}
\label{figS:matchedslices}
\end{figure}

\begin{figure}[ht]
\centering
\includegraphics[width=0.76\textwidth]{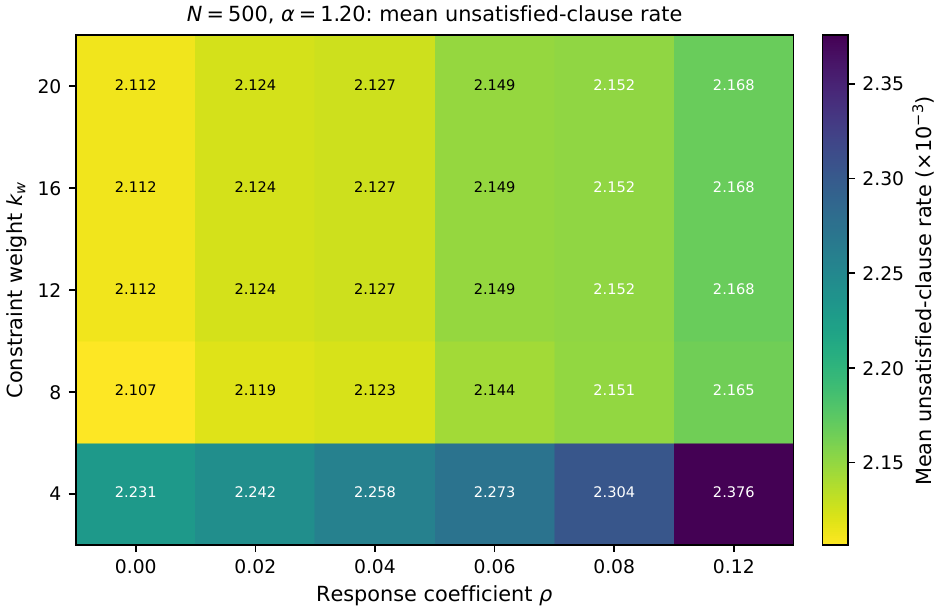}
\caption{Matched-$k_w$ random 2-SAT response surface.  The memoryless $\rho=0$ edge contains the optimum for every clause weight.}
\label{figS:matchedheatmap}
\end{figure}

\begin{table}[ht]
\caption{Mean selected unsatisfied-clause rate in the matched-$k_w$ response surface.}
\label{tabS:2satgrid}
\begin{ruledtabular}
\begin{tabular}{c@{\quad}cccccc}
$k_w\backslash\rho$ & 0 & 0.02 & 0.04 & 0.06 & 0.08 & 0.12 \\
\hline
4  & 0.0022311 & 0.0022422 & 0.0022578 & 0.0022733 & 0.0023039 & 0.0023756 \\
8  & 0.0021067 & 0.0021194 & 0.0021233 & 0.0021439 & 0.0021506 & 0.0021650 \\
12 & 0.0021122 & 0.0021244 & 0.0021267 & 0.0021494 & 0.0021517 & 0.0021678 \\
16 & 0.0021122 & 0.0021244 & 0.0021267 & 0.0021494 & 0.0021517 & 0.0021678 \\
20 & 0.0021122 & 0.0021244 & 0.0021267 & 0.0021494 & 0.0021517 & 0.0021678 \\
\end{tabular}
\end{ruledtabular}
\end{table}

The predeclared primary contrast is $k_w=16$, $\rho=0.08$ versus 0.  Mean rate changes from 0.00211222 to 0.00215167: treatment minus control is $+3.944\times10^{-5}$ with formula-bootstrap interval $[1.333,6.611]\times10^{-5}$.  This equals 0.02367 additional unsatisfied clauses per trial or a 1.867\% worsening.  Formula-wise counts are 8 improved, 17 worsened, and 5 tied; paired standardized effect $d_z=0.519$.  Terminal unsatisfied rate worsens by 11.35\%.

SAT success changes by $-1.46$ percentage points with interval $[-3.77,+0.46]$ percentage points.  Among unsatisfiable formulas, selected unsatisfied-clause rate increases by $4.804\times10^{-5}$ with interval $[0.882,8.627]\times10^{-5}$.  Neither stratum provides positive response-memory evidence.  Because the predeclared primary test did not support a positive effect, the prespecified lower-density extension was not performed, and no $\alpha$ dependence is claimed.

\section{Validation and invariant tests}
\label{secS:validation}

The following invariants were checked before statistical analysis and inclusion in the reported results:
\begin{enumerate}
\item At $\lambda=0$, \psa{}, additive, normalized, and gain-only kernels produce identical trajectories for shared random streams and matched parameters.
\item At $\rho=0$, finite-response dynamics are trajectory-identical to \psa{} with matched parameters.
\item At $\kappa=0$, binary-state self-feedback is trajectory-identical to matched \psa{} under common random streams; the implementation uses the pre-update/current bit state and allocates no separate response state.
\item Independent additive implementations agree throughout the common evaluated coefficient range.
\item Logged trajectory endpoints reproduce the high-statistics control implementation outputs bit for bit for all nonshuffled variants.
\item The shuffled mapping is a derangement, has no self-links, and preserves delayed-response mean and second moment to numerical precision.
\item The arbitrary-start event kernel exactly reproduces the existing all-zero event kernel when supplied the all-zero state.  For alternative starts, bit-state, channel, and transmitted-word hashes match across the four paired methods, and every response state begins at zero.
\item All cross-code matrices satisfy requested row/column degrees, GF(2) rank checks, rate checks, and stored SHA-256 identities.  The pSA-specific dictionaries match the independently optimized representative-code selections, have zero memory coefficients, and remain fixed across all ten matrices at each size.
\item The binary-state high-statistics records reproduce the selected and $N=192$ sensitivity configurations, and the score/BER/FER selection audit agrees at $N=96$ and 288 while preserving the prespecified $N=192$ score selection separately.
\item The matched-$k_w$ 2-SAT control uses the same formula list and paired seed streams in every $\rho$ slice at fixed $k_w$; formula identities and satisfiability labels were verified.
\item Reported cross-code intervals resample codes, initialization-acquisition intervals use ten paired seed-batch means, and matched-$k_w$ 2-SAT intervals resample formulas.  None treats bits as independent instances for the corresponding claim.
\end{enumerate}

For implementation provenance, the optimized LDPC Numba path for the memoryless, additive, normalized-memory, and gain-only rules samples holding first and skips held candidates, whereas vectorized finite-response LDPC and cross-problem paths form candidates before masking.  For a fixed rule, the two orderings are transition-law equivalent but consume pseudorandom numbers differently, so equal seeds across paths need not yield identical trajectories.  The public release documents these paths and provides scripts, dependencies, parameters, seeds, metadata, checksums, binary-state-control records, three-arm transfer records, initialization-validation records, and reproduction commands \cite{onizawa_psa_ldpc_zenodo_2026}.

\bibliography{references}